\documentclass[11pt]{article}
\usepackage{tabularx}
\usepackage{array}
\usepackage{graphics}
\usepackage{graphicx}
\usepackage{epsfig}
\usepackage{amsmath}
\usepackage{amssymb}
\usepackage{axodraw}
\makeatletter

\usepackage{verbatim}

\newcommand{\bea}{\begin{eqnarray}}
\newcommand{\eea}{\end{eqnarray}}
\def\beq{\begin{equation}}
\def\eeq{\end{equation}}

\newcommand{\pdir}{p\kern -5.2pt\raise 0.2ex\hbox {/}}
\newcommand{\vdir}{v\kern -5.75pt\raise 0.15ex\hbox {/}}
\newcommand{\kdir}{k\kern -5.75pt\raise 0.15ex\hbox {/}}
\newcommand{\epsdir}{\epsilon\kern -5.0pt\raise 0.15ex\hbox {/}}
\newcommand{\bvdir}{\bar{v}\kern -5.75pt\raise 0.15ex\hbox {/}}
\newcommand{\Ddir}{D\kern -7.75pt\raise 0.20ex\hbox {/}}
\newcommand{\Adir}{A\kern -7.75pt\raise 0.20ex\hbox {/}}
\newcommand{\ldir}{l\kern -5.0pt\raise 0.2ex\hbox{/}}
\newcommand{\varepsdir}{\varepsilon\kern -5.5pt\raise 0.15ex\hbox{/}}

\newcommand{\pslash}{p \!\!\!/}

\makeatother

\begin{document}
\thispagestyle{empty} 

\begin{flushright}
\begin{tabular}{l}
{\tt LPT Orsay 02-99}\\
{\tt UHU-FT/03-12 }\\
{\tt FAMN SE-01/03}\\
{\tt CPHT RR 032.0603}\\
\end{tabular}
\end{flushright}
\vskip 2.2cm\par
\begin{center}
%\today\hskip 1 cm
%version 3.7

%{\par\centering \textbf{\LARGE Systematic correction of hypercubic } }\\
%\vskip .45cm\par
%{\par\centering \textbf{\LARGE artifacts from lattice simulations} }\\
{\par\centering \textbf{\LARGE
Quark propagator and vertex: systematic corrections of hypercubic
artifacts from lattice simulations}}

\vskip 0.9cm\par
{\par\centering \large  
\sc  Ph.~Boucaud$^a$,  F. de Soto$^b$
J.P.~Leroy$^a$, A.~Le~Yaouanc$^a$, J. Micheli$^a$, 
H.~Moutarde$^c$, O.~P\`ene$^a$, J.~Rodr\'{\i}guez--Quintero$^d$ }
{\par\centering \vskip 0.5 cm\par}
{\par\centering \textsl{ 
$^a$~Laboratoire de Physique Th\'eorique (B\^at.210), Universit\'e de
Paris XI,\\ 
Centre d'Orsay, 91405 Orsay-Cedex, France.} \\
\vskip 0.3cm\par }{\par\centering \textsl{  
$^b$ Dpto. de F\'{\i}sica At\'omica, Molecular y Nuclear \\
Universidad de Sevilla, Apdo. 1065, 41080 Sevilla, Spain }\\
\vskip 0.3cm\par }{\par\centering \textsl{ 
$^c$ Centre de Physique Th\'eorique Ecole Polytechnique, 
91128 Palaiseau Cedex, France }\\
\vskip 0.3cm\par }{\par\centering \textsl{ 
 $^d$ Dpto. de F\'{\i}sica Aplicada e Ingenier\'{\i}a el\'ectrica \\
E.P.S. La R\'abida, Universidad de Huelva, 21819 Palos de la fra., Spain} 
\vskip 0.3cm\par }

  \end{center}

\vskip 0.45cm
\begin{abstract}
This is the first part of a study of the quark propagator and 
the  vertex function of the vector
current  on the lattice in the Landau gauge  and using both Wilson-clover and
overlap actions.  In order to be able to identify lattice artifacts and 
to reach large momenta we  use a range of lattice spacings. The lattice
artifacts turn out to be exceedingly large in this  study. We present a
new and very efficient  
 method to eliminate the hypercubic (anisotropy) artifacts based on a
systematic expansion on hypercubic invariants which are not $SO(4)$
invariant. A simpler version of this method has been used in previous works.
This method is shown to be significantly more efficient than
the popular ``democratic'' methods. It can of course be applied to the 
lattice simulations of many other physical quantities. 
The analysis indicates a hierarchy in the size of hypercubic artifacts:
 overlap  larger than clover  and  propagator larger than  vertex function.  
This pleads for the combined study of propagators and vertex functions
via  Ward identities. 
\end{abstract}
\vskip 0.4cm
{\small PACS: \sf 12.38.Gc  (Lattice QCD calculations)}            
\vskip 2.2 cm 
 
\setcounter{page}{1}
\setcounter{footnote}{0}
\setcounter{equation}{0}
%%%%%%%%%%%%%%%%%%%%%%%%%%%%%%%%%%%%%%%%
%%%%%%%%%%%%%%%%%%%%%%%%%%%%%%%%%%%%%%%%
%%%%%%%%%%%%%%%%%%%%%%%%%%%%%%%%%%%%%%%%

\renewcommand{\thefootnote}{\arabic{footnote}}
\vspace*{-1.5cm}
\newpage
\setcounter{footnote}{0}
%%%%%%%%%%%  Section 1
\section{Introduction}

The study of the quark propagator and vertex functions has been 
extensively pursued in the literature starting in the 70's 
\cite{lane}. Lattice QCD has more recently  treated this issue
\cite{pittori2}. A systematic treatment varying the quark actions  has
been followed by the CSSM collaboration~\cite{adelaide}. The scalar
part of the quark propagator  is related  via Ward identities to the
pseudoscalar vertex function. The role of the Goldstone boson pole in
the latter has been  thoroughly discussed \cite{Cudell:2001ny}.

Leaving aside the latter issue, we will mainly concentrate on the 
vector part of the quark propagator, the one which is proportional to
$\pslash $.  One of our main goals is to check the effect of the $A^2$
condensate  which has been discovered via power corrections at large
momenta  to the gluon propagator and three point Green 
functions~\cite{Boucaud:2000ey}-\cite{Becirevic:1999hj}. The values
plotted in literature for $Z_\psi(p^2)$ are extremely flat above 2
GeV~\cite{Becirevic:1999kb}. At  first sight this is a satisfactory
feature since the perturbative-QCD corrections  are known to be small.
However a closer scrutiny   makes it worrying since both the
perturbative QCD corrections and the $A^2$ condensate  predict a
decrease which seems not to be seen. This leads us to start a very
systematic study of the   problem, with the following series of
improvements  on  earlier works:

\begin{itemize}
\item We reach an energy of 10~GeV by matching several lattice spacings 
so we are in  a better position to eliminate lattice artifacts.

\item We make a systematic use of Ward identities relating the quark propagator
and the  vertex function  via the constant $Z_V$ and study both quantities in 
parallel.

\item We make use of a very efficient way of eliminating hypercubic
artifacts. A simpler version of it was elaborated  while studying  gluon
propagators~\cite{Boucaud:1998bq},~\cite{Becirevic:1999uc}.  In this
work we have encountered the necessity to improve significantly this method.

\end{itemize}

\begin{figure}[hbt]
\begin{center}
\leavevmode
\mbox{\epsfig{file=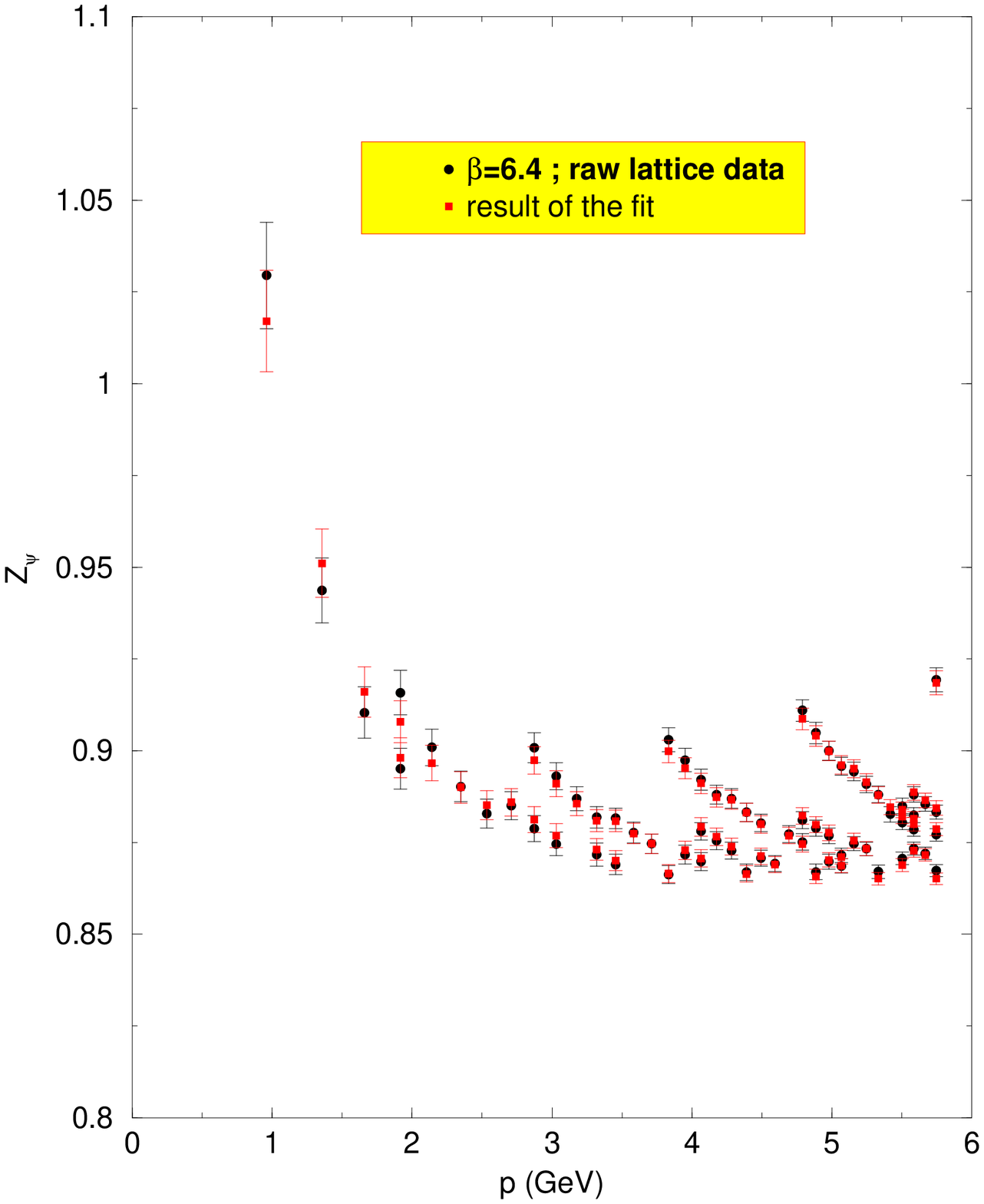,height=8cm}}
\mbox{\epsfig{file=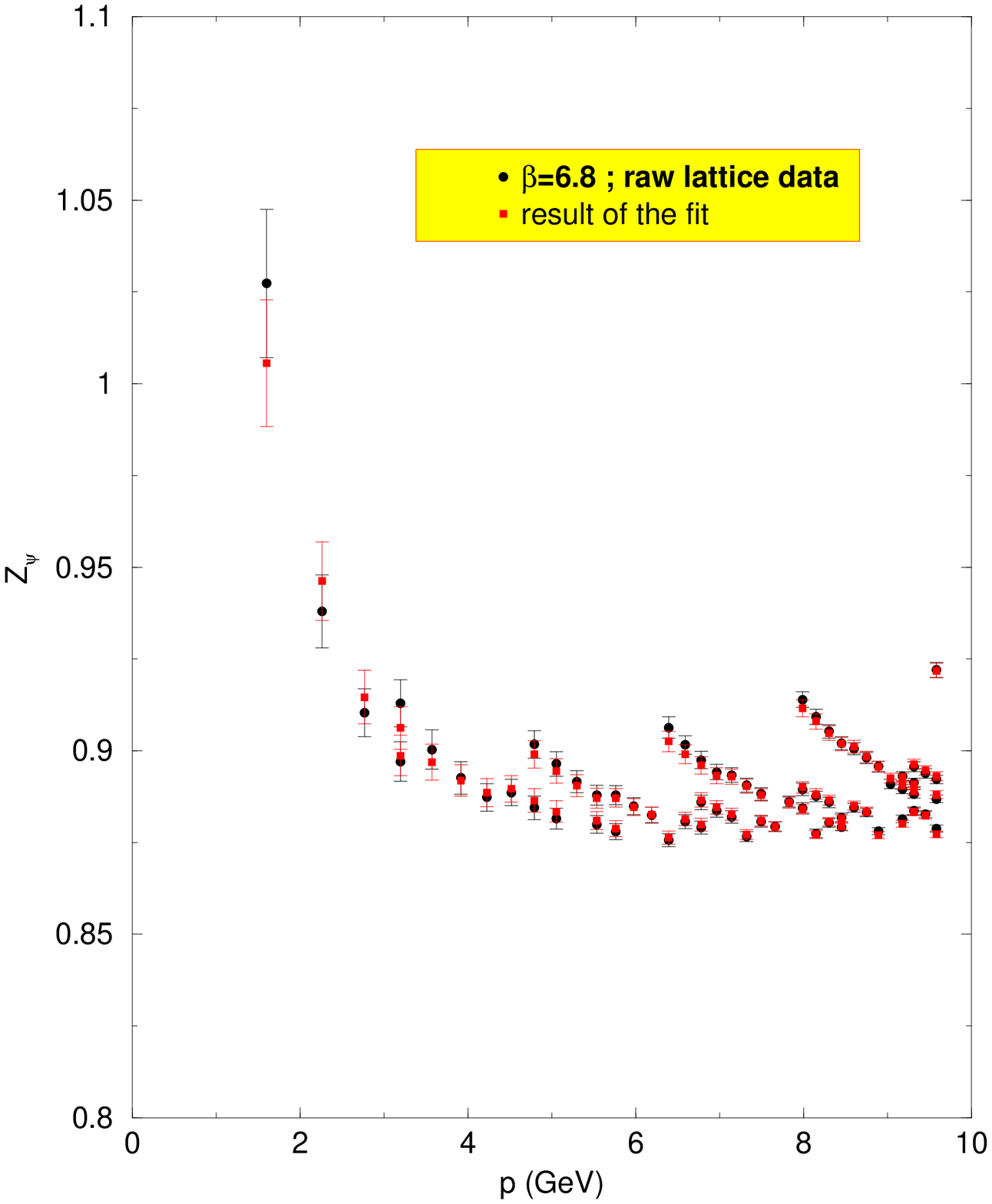,height=8cm}}
%\vskip -2.5 cm
\caption{\small Fit of the raw lattice data for clover-$Z_\psi$
with formula (\ref{micheli2-1}) for $\beta=6.4$ and 6.8. 
For $Q(p^2, 0, 0)$ we have taken a polynomial of degree eight 
in $p^2$. The fit has been done independently for each $\beta$.
 The raw data
are represented by black circles while the fit is shown by 
red squares. The agreement is often such that the raw data and
fit points are indistinguishable on the plot.
  }
\label{FigH4}
\end{center}
\end{figure}

This last point will be the main subject of this paper. 
Indeed, the raw data show a shape somewhat reminiscent 
of a half-fishbone (fig.~\ref{FigH4}), utterly different from a smooth curve
expected in the continuum. 
As we shall see the method elaborated 
in~\cite{Boucaud:1998bq},~\cite{Becirevic:1999uc} proves not
to be powerful enough. We therefore wish to attract attention 
 on the generalisation of the above-mentioned method 
 which we believe is strikingly efficient and should become 
 a very useful tool for the lattice community. 

The remaining part of the work, i.e. the correction of 
$SO(4)$ symmetric artifacts and the resulting physics results
will be presented in a later publication. 

In  section \ref{theor} we will recall some theoretical premises, 
in section \ref{lattice} we will indicate the lattice simulations
 which we have performed, in section \ref{arti}  we will describe our 
 method to eliminate lattice artifacts and compare it to earlier methods.

%============================================
%============================================
\section{Theoretical premisses}
\label{theor}

We work in the Landau gauge. Let us first fix the notations that we will use.
We will use all along the Euclidean metric. The continuum quark propagator
is a $12 \times 12$ matrix $S(p_\mu)$. The inverse propagator is expanded
\bea\label{Sm1}
\widetilde S^{-1}(p) = \delta_{a,b} Z_\psi(p^2) 
\left( i\,\pslash + m(p^2)\right)
\eea
where $a,b$ are the color indices. 

Let us consider a colorless vector current $\bar q \gamma_\mu q$.
 The three point Green function  $G_\mu$ is defined  by
\bea\label{Gmu}
G_\mu(p, q) = \int d^4x d^4y e^{i p \cdot y - i (p+q) \cdot x} 
< q(y)\bar q(0) \gamma_\mu q(0) \bar q(x) >\ .
\eea
In all this paper we will restrict ourselves to the case where
the vector current carries a vanishing momentum transfer $q_\mu$.
The vertex function is then defined by 
\bea\label{gammamu1}
\Gamma_\mu(p, q=0) = \widetilde S^{-1}(p) \,G_\mu(p, q=0)\,
\widetilde S^{-1}(p)
\ .
\eea
In the following we will omit to write $q_\mu=0$ and we will understand
$\Gamma_\mu(p)$ as the bare vertex function computed on the lattice.
The renormalised vertex function is $Z_V\Gamma_\mu(p)$. 

~From Lorentz covariance and discrete symmetries
\bea\label{gammamu2}
\Gamma_\mu(p) = \delta_{a,b} \left[g_1(p^2) \gamma_\mu + i g_2(p^2) p_\mu +
g_3(p^2) p_\mu \pslash + i g_4(p^2) [\gamma_\mu,\pslash]\right]\ .
\eea

The Ward identity tells us that
\bea\label{ward1}
Z_V\Gamma_\mu(p) = -i\, \frac {\partial}{\partial p^\mu}
\widetilde  S^{-1}(p),\quad
\eea
which from (\ref{Sm1})-(\ref{gammamu2}) implies
\bea\label{ward2}
Z_\psi(p^2) = Z_V g_1(p^2),\quad 
2\,\frac {\partial}{\partial p^2}Z_\psi(p^2)=Z_V g_3(p^2),\nonumber \\ 
2\,\frac {\partial}{\partial p^2}b(p^2)=- Z_V g_2(p^2),\quad g_4(p^2)=0.
\eea

For a conserved current, $Z_V=1$. We keep $Z_V$ since the local  vector
current on the lattice is not conserved; it will differ from 1 by
lattice perturbative corrections which are a finite  series in the
``boosted'' bare coupling constant,  independent of $p^2$. However,
lattice artifacts do generate a sometimes significant $p^2$ dependence 
of $Z_V$ at the level of raw lattice data, see for example
fig.~\ref{Fig:Zv-over}. We will therefore  sometimes use a p-dependent
raw $Z_V$ written $Z_V(p^2)$ and defined  in eq. (\ref{defzv})

The renormalisation scheme that we use is the one called 
MOM', eq (26) in ref. \cite{Chetyrkin:1999pq}. The bare propagator 
$S(p)$ is multiplied by the renormalisation constant 
\bea
\widetilde S_{\rm R}(p) = Z_\psi(\mu)\widetilde  S(p),\quad 
{\rm whence}
\quad \widetilde S_{\rm R}^{-1}(p)|_{p^2=\mu^2} =
 \delta_{a,b} \left( i\,\pslash + m(p^2)\right)
\eea
This defines the quark field renormalisation. Due to the Ward identity, 
the factor $Z_\psi(\mu)^{-1}$ multiplies the bare vertex function 
$g_1$, so that $g_1^{\rm R}(p^2=\mu^2)= Z_V^{-1}$.

~From the anomalous dimensions computed in ref. \cite{Chetyrkin:1999pq} 
we may express the perturbative running of $Z_\psi$ for example as a 
function of the running $\alpha_{\overline {\rm MOM}}(p)$. As $Z_V$ in
the continuum is a constant, $Z_\psi(p)$ and $g_1(p)$ have the same 
perturbative scale dependence.

%%%%%%%%%%%%%%%%%%%%%%%%%%%%%%%%%%%%%%%%%%%%%%%%%%%%%%%%%%%%%%%%%%%%%%
%%%%%%%Figure 1
%%%%%%%%%%%%%%%%%%%%%%%%%%%%%%%%%%%%%%%%%%%%%%%%%%%%%%%%%%%%%%%%%%%%%%
\vskip 1cm
%%\vspace*{-4.5cm}
%%%%%%%%%%%%%%%%%%%%%%%%%%%%%%%%

%%%%%%%Figure 2
%%%%%%%%%%%%%%%%%%%%%%%%%%%%%%%%%%%%%%%%%%%%%%%%%%%%%%%%%%%%%%%%%%%%%%
\begin{figure}[t]
\begin{center}
\vskip 2cm
\leavevmode
\vskip -2.5 cm
%\mbox{\epsfig{file=BB_orb_6.0.eps,height=8cm}}
%\mbox{\epsfig{file=BB_orb_6.8.eps,height=8cm}}
\mbox{\epsfig{file=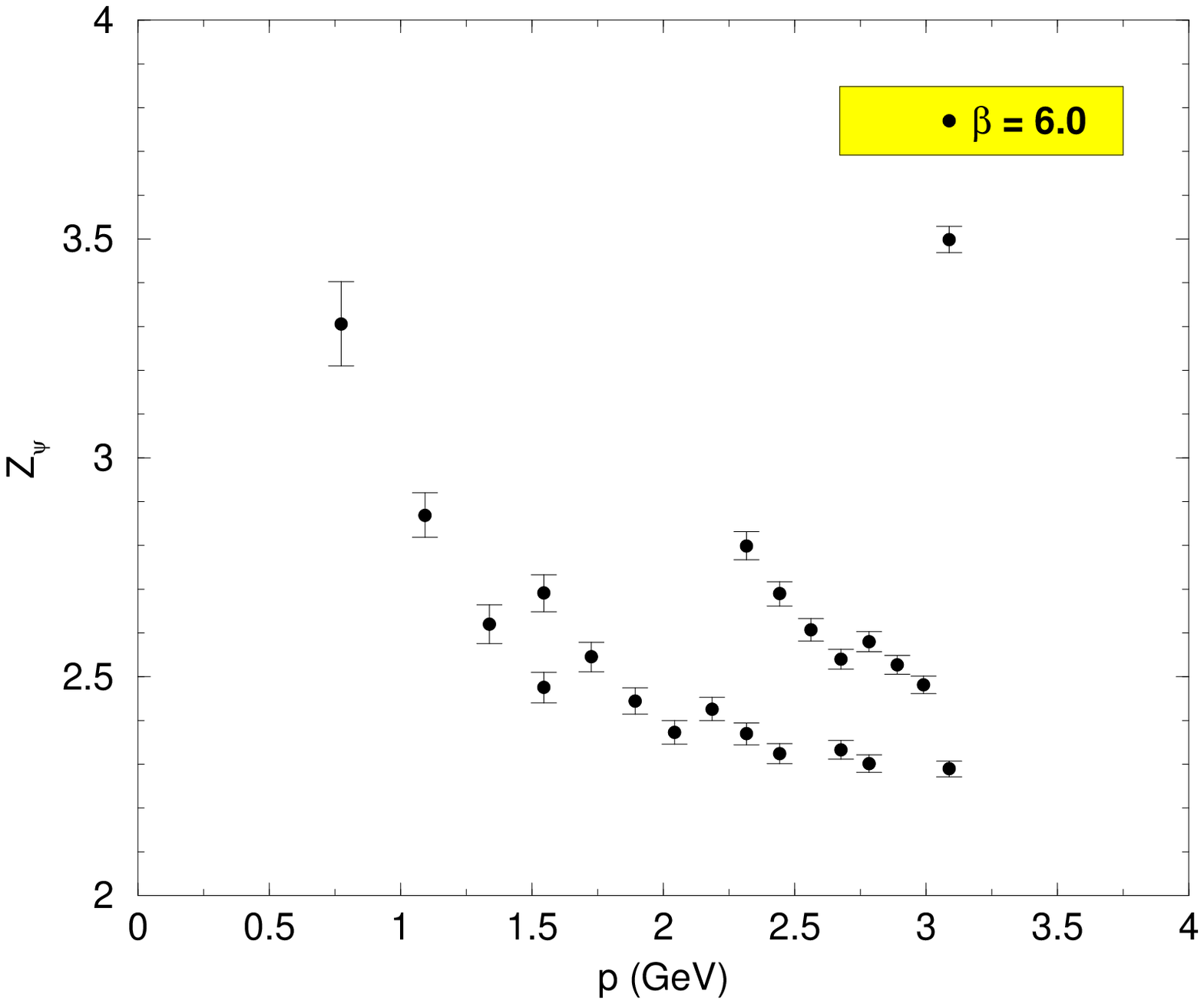,height=6cm}}
\mbox{\epsfig{file=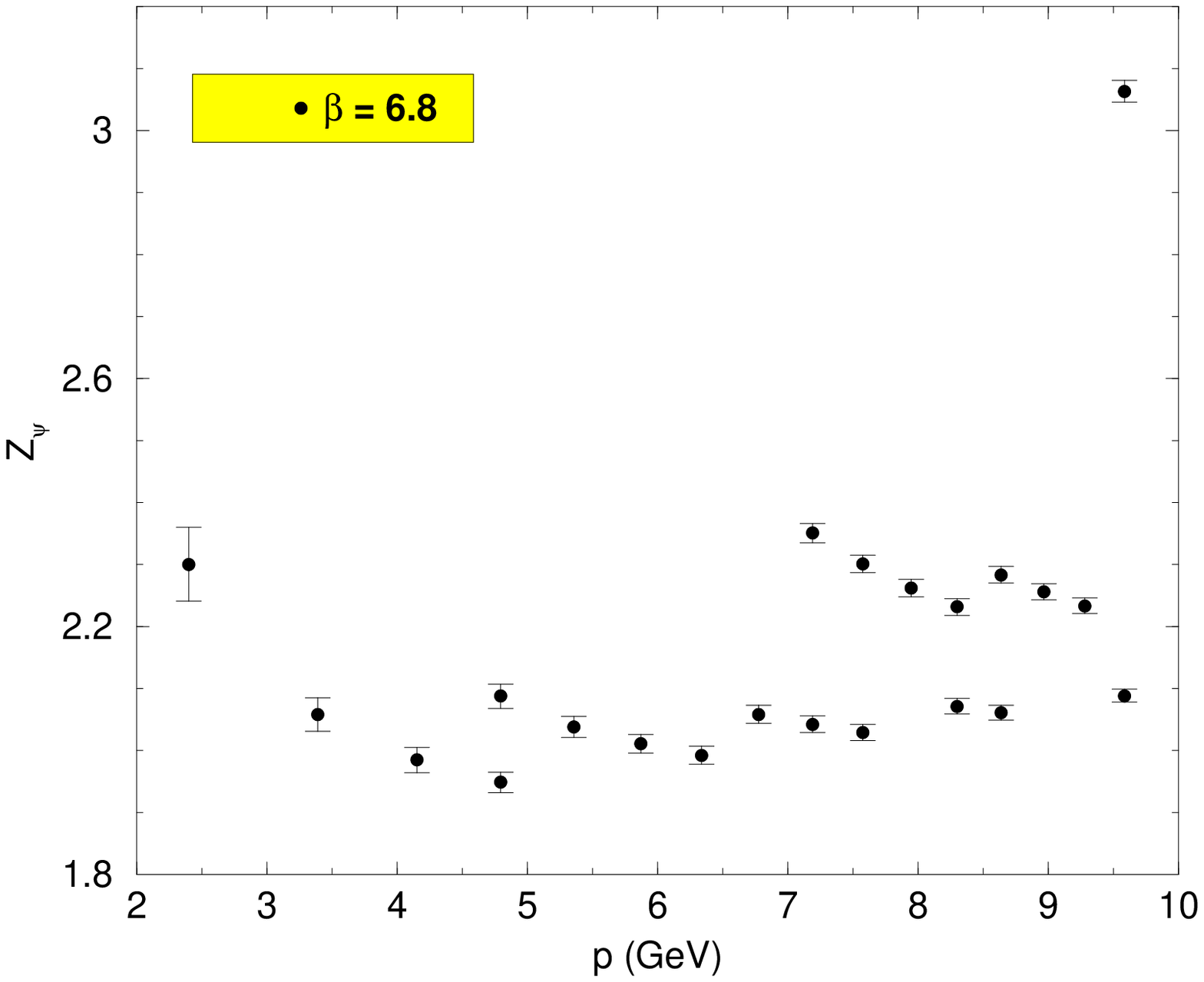,height=6cm}}
\caption{\small Raw lattice data for overlap-$Z_\psi$ 
 for $\beta=6.0$ and 6.8.}
\label{Fig:Zpsi-over}
\end{center}
\end{figure} 
%%\vspace*{-4.5cm}
%%%%%%%%%%%%%%%%%%%%%%%%%%%%%%%%

%%%%%%%Figure 3
%%%%%%%%%%%%%%%%%%%%%%%%%%%%%%%%%%%%%%%%%%%%%%%%%%%%%%%%%%%%%%%%%%%%%%
\begin{figure}[t]
\begin{center}
\vskip 2cm
\leavevmode
\vskip -2.5 cm
%\mbox{\epsfig{file=zv_orb_6.0.eps,height=8cm}}
%\mbox{\epsfig{file=zv_orb_6.8.eps,height=8cm}}
\mbox{\epsfig{file=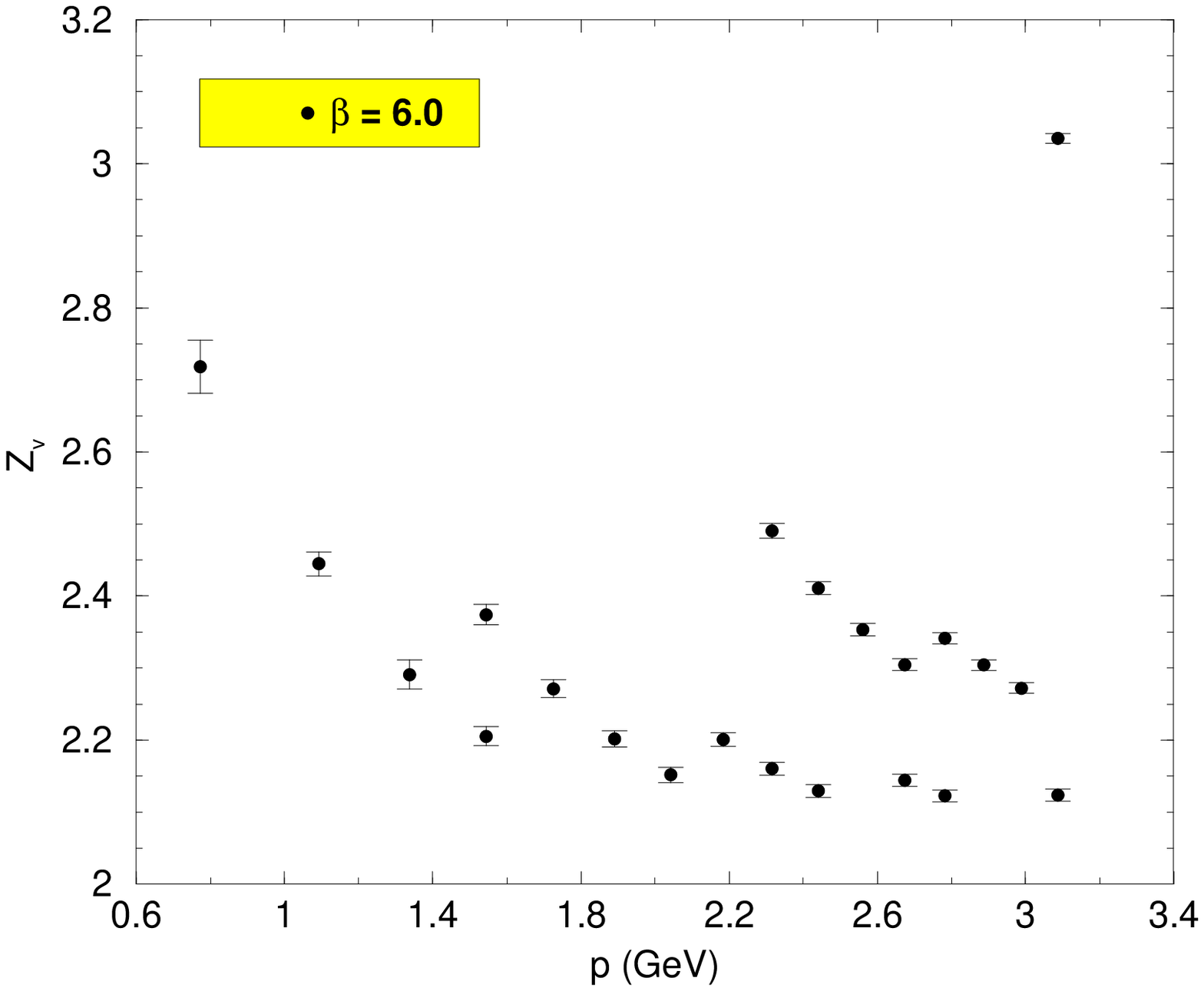,height=6cm}}
\mbox{\epsfig{file=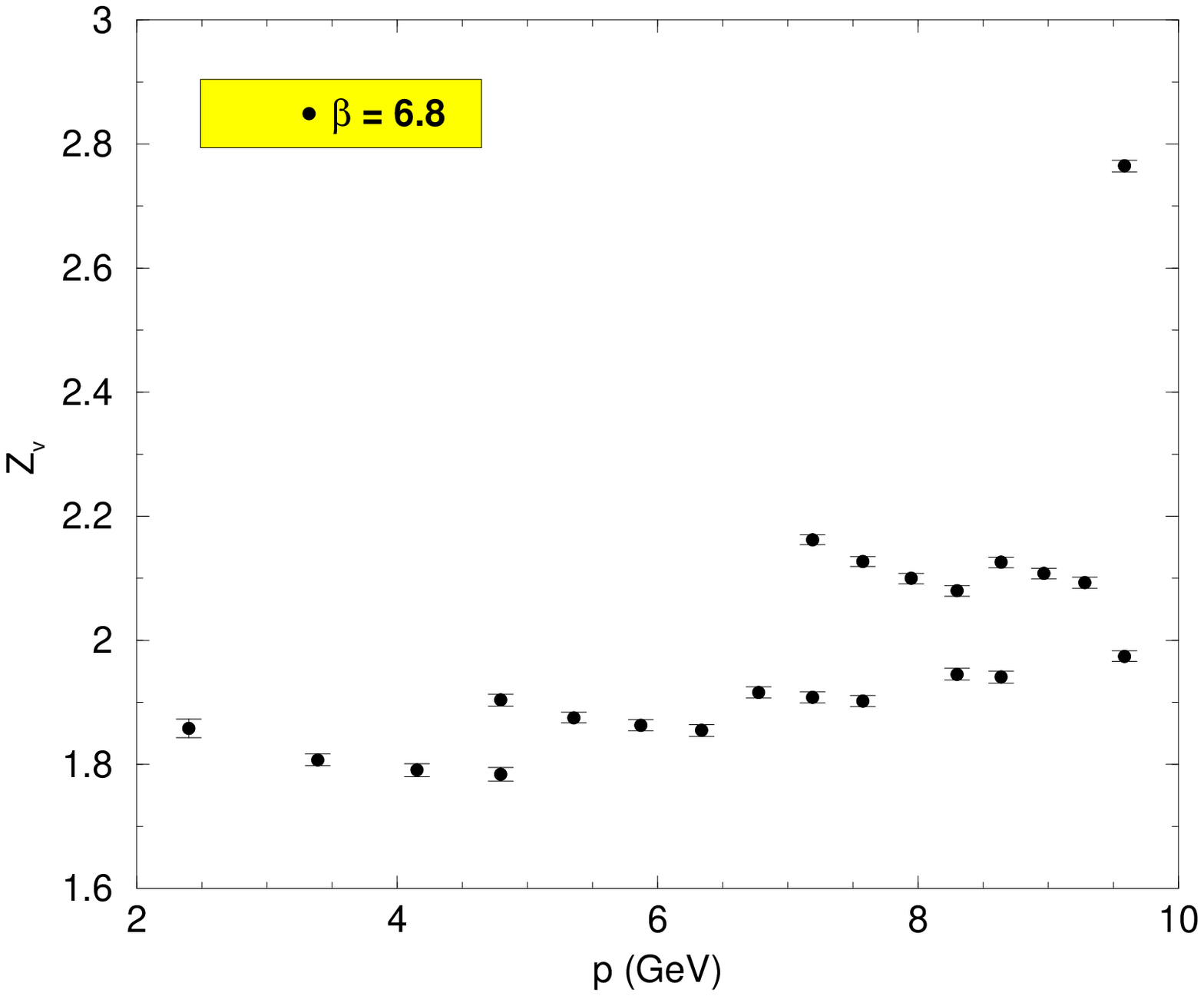,height=6cm}}
\caption{\small Raw lattice data for overlap-$Z_V$ 
 for $\beta=6.0$ and 6.8.}
\label{Fig:Zv-over}
\end{center}
\end{figure} 
%%\vspace*{-4.5cm}
%%%%%%%%%%%%%%%%%%%%%%%%%%%%%%%%

%%%%%%%%%%%%%%%%%%%%%%%%%%%%%%%%%%%%%%%%%%%%%%%%%%%%%%%%%%%%%%%%%%%%%%
%%%%%%%Figure 4
%%%%%%%%%%%%%%%%%%%%%%%%%%%%%%%%%%%%%%%%%%%%%%%%%%%%%%%%%%%%%%%%%%%%%%
\begin{figure}[hbt]
\begin{center}
\leavevmode
%\vskip 1.5 cm
%\mbox{\epsfig{file=comp-R-M-6.4.eps,height=8cm}}
%\mbox{\epsfig{file=comp-R-M-6.8.eps,height=8cm}}
\mbox{\epsfig{file=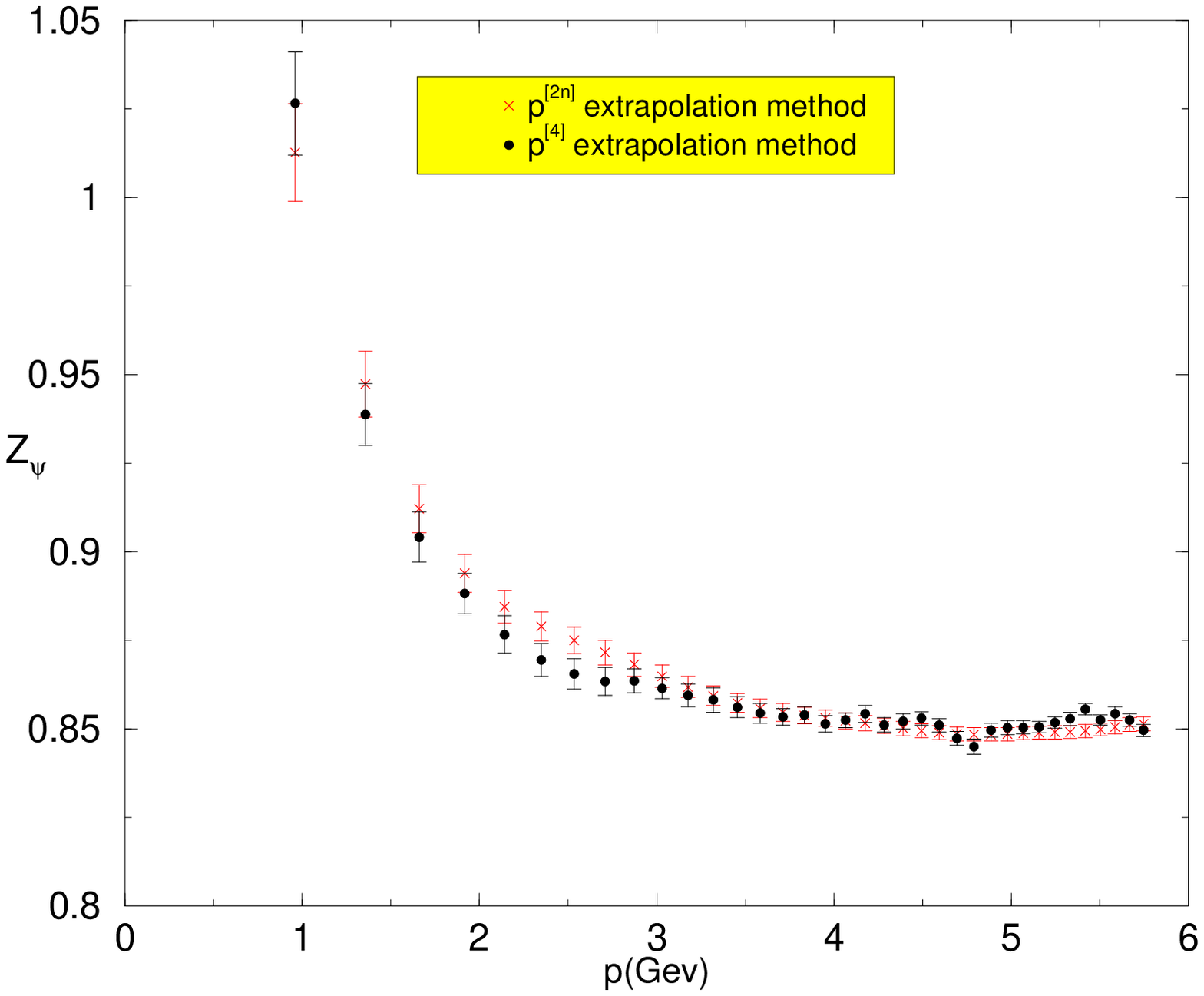,height=6cm}}
\mbox{\epsfig{file=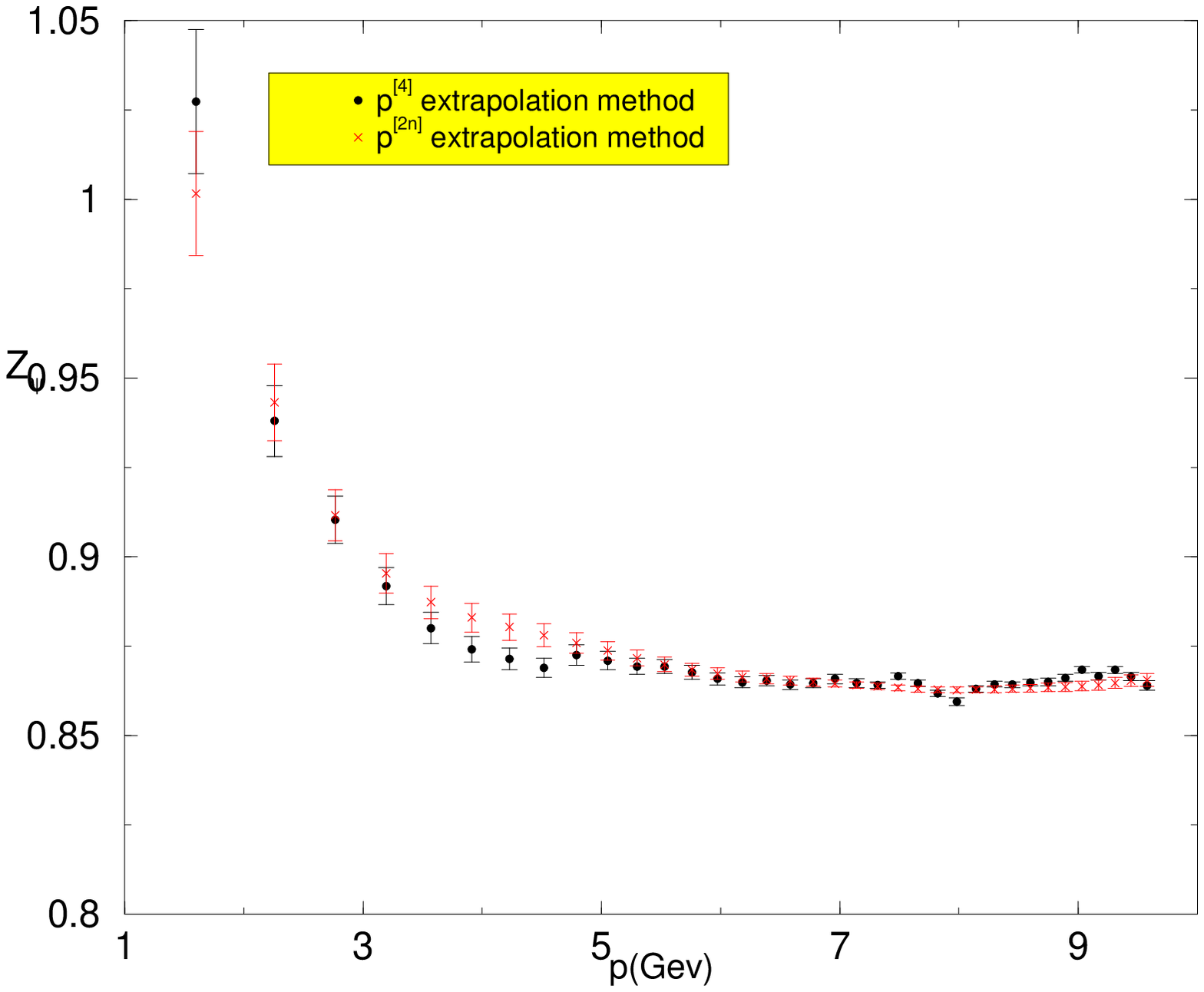,height=6cm}}
\caption{\small comparison of the  ``$\mathbf p^{[4]}$ 
extrapolation method, represented by black circles,
 with the  `` $\mathbf p^{[2n]}$ 
extrapolation method '' represented by ``x'' symbols.
The left (right) plot shows the result for clover-$Z_\psi$ 
at $\beta=6.4$ ($\beta=6.8$)}
\label{FigRM}
\end{center}
\end{figure} 
%%\vspace*{-4.5cm}
%%%%%%%%%%%%%%%%%%%%%%%%%%%%%%%%

%%%%%%%Figure 5
%%%%%%%%%%%%%%%%%%%%%%%%%%%%%%%%%%%%%%%%%%%%%%%%%%%%%%%%%%%%%%%%%%%%%%
\begin{figure}[t]
\begin{center}
\leavevmode
%\mbox{\epsfig{file=BB_6.0_R.eps,height=8cm}}
%\mbox{\epsfig{file=BB_6.8_R.eps,height=8cm}}
\mbox{\epsfig{file=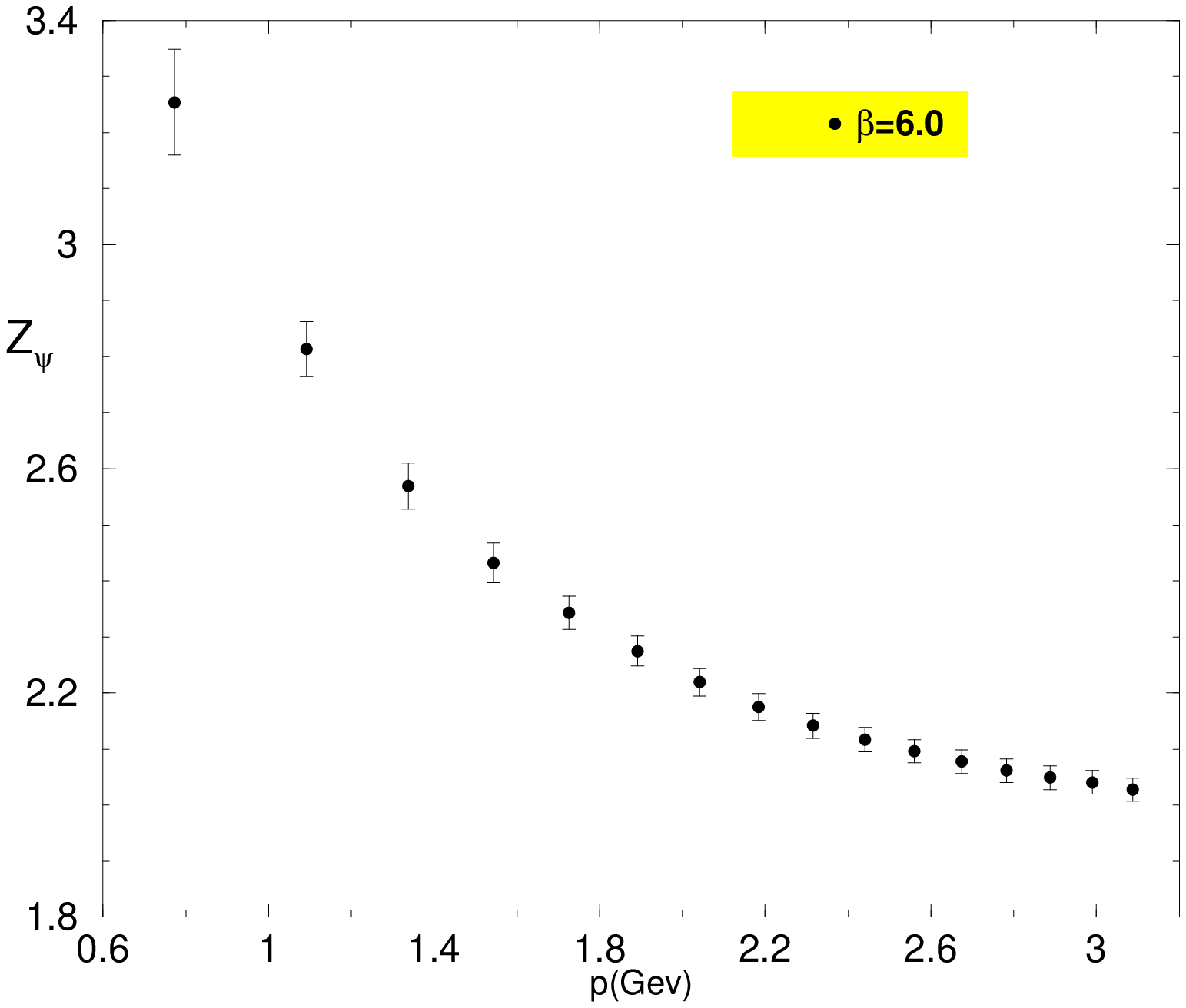,height=6cm}}
\mbox{\epsfig{file=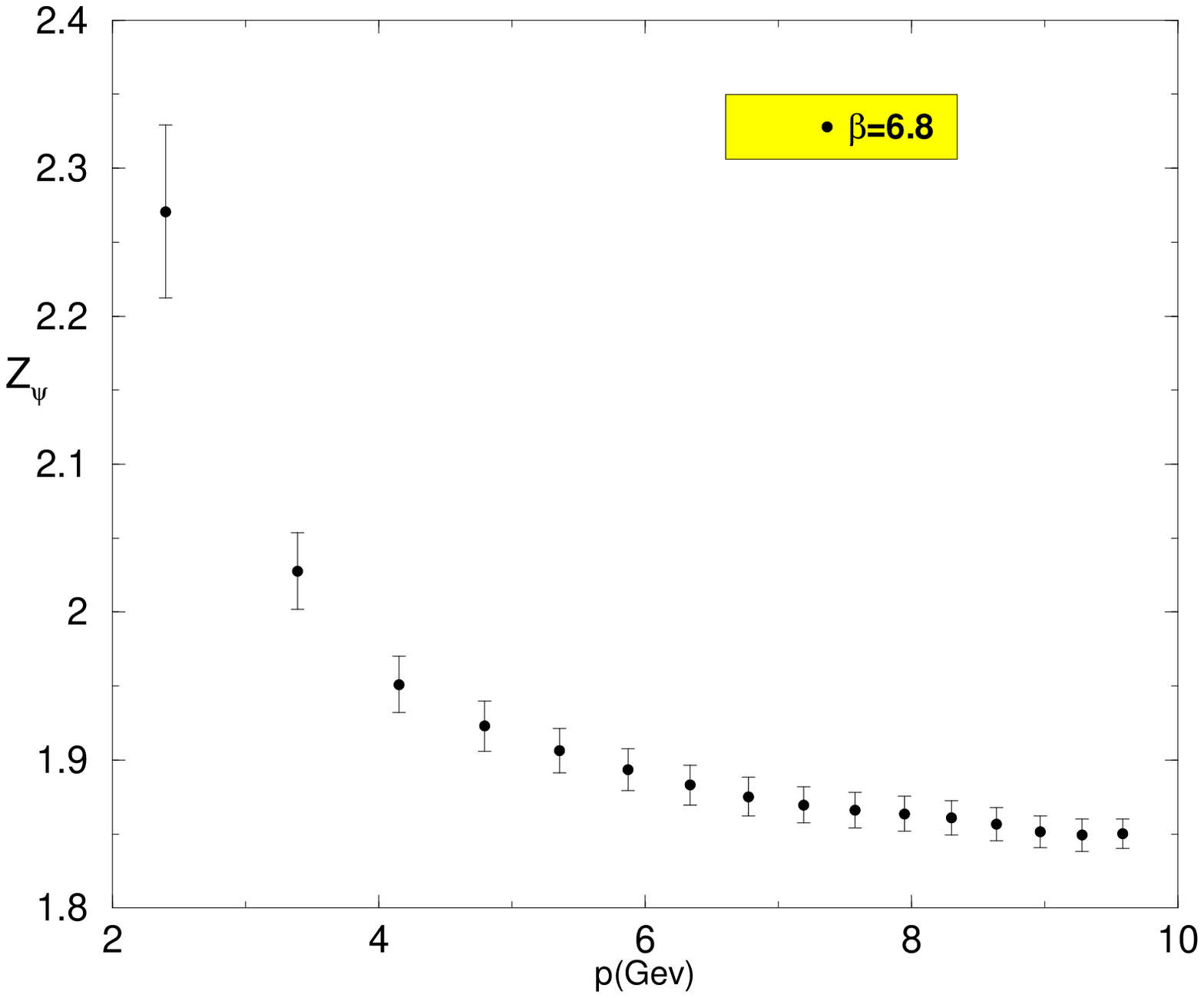,height=6cm}}
\caption{\small Lattice data for overlap-$Z_\psi$ 
after application of the `` $\mathbf p^{[2n]}$ 
extrapolation method '' for $\beta=6.0$ and 6.8.}
\label{Fig:Zpsi-over-R}
\end{center}
\end{figure} 
%%\vspace*{-4.5cm}
%%%%%%%%%%%%%%%%%%%%%%%%%%%%%%%%

%%%%%%%Figure 6
%%%%%%%%%%%%%%%%%%%%%%%%%%%%%%%%%%%%%%%%%%%%%%%%%%%%%%%%%%%%%%%%%%%%%%
\begin{figure}[t]
\begin{center}
\leavevmode
%\mbox{\epsfig{file=zv_6.0_R.eps,height=8cm}}
%\mbox{\epsfig{file=zv_6.8_R.eps,height=8cm}}
\mbox{\epsfig{file=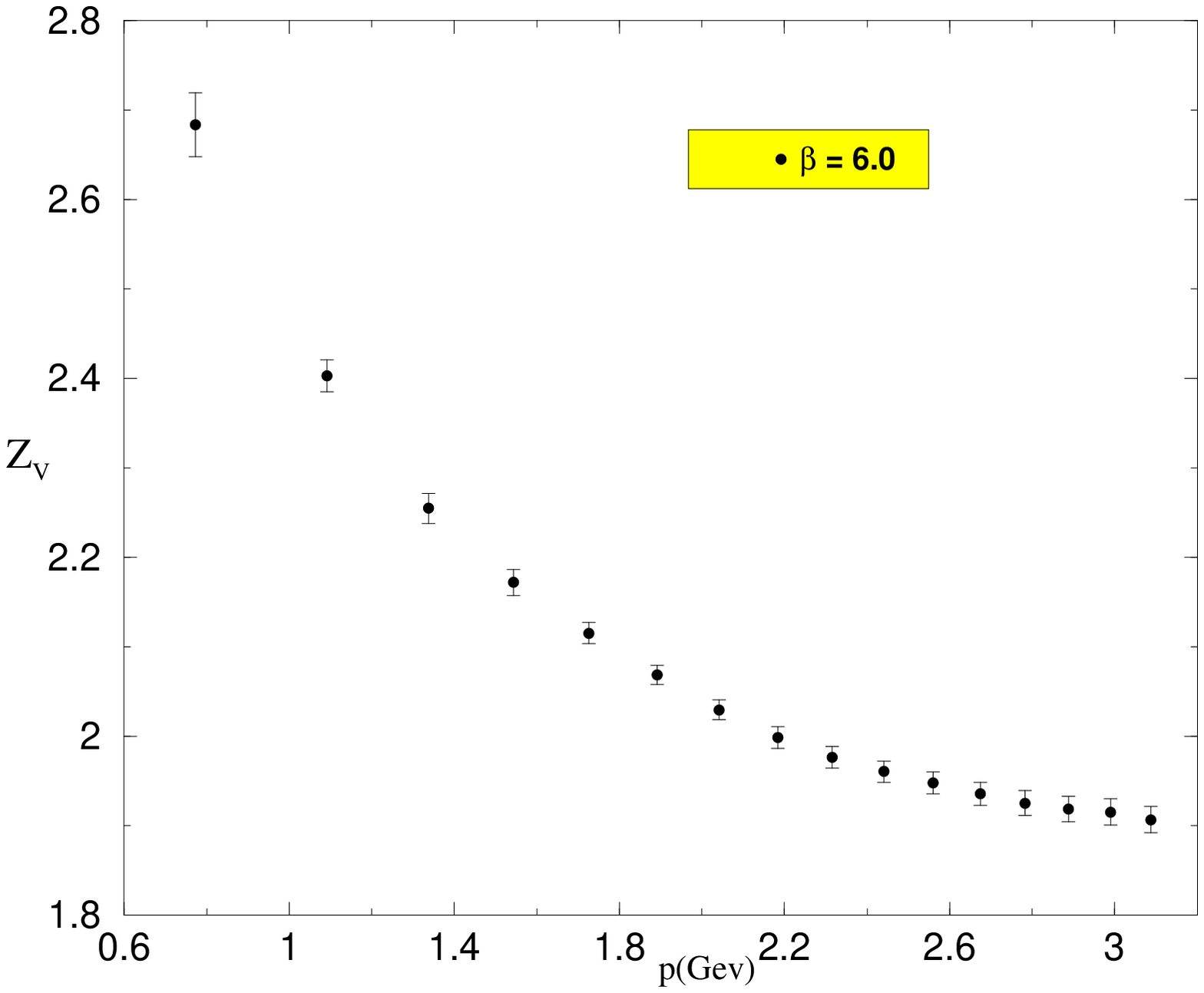,height=6cm}}
\mbox{\epsfig{file=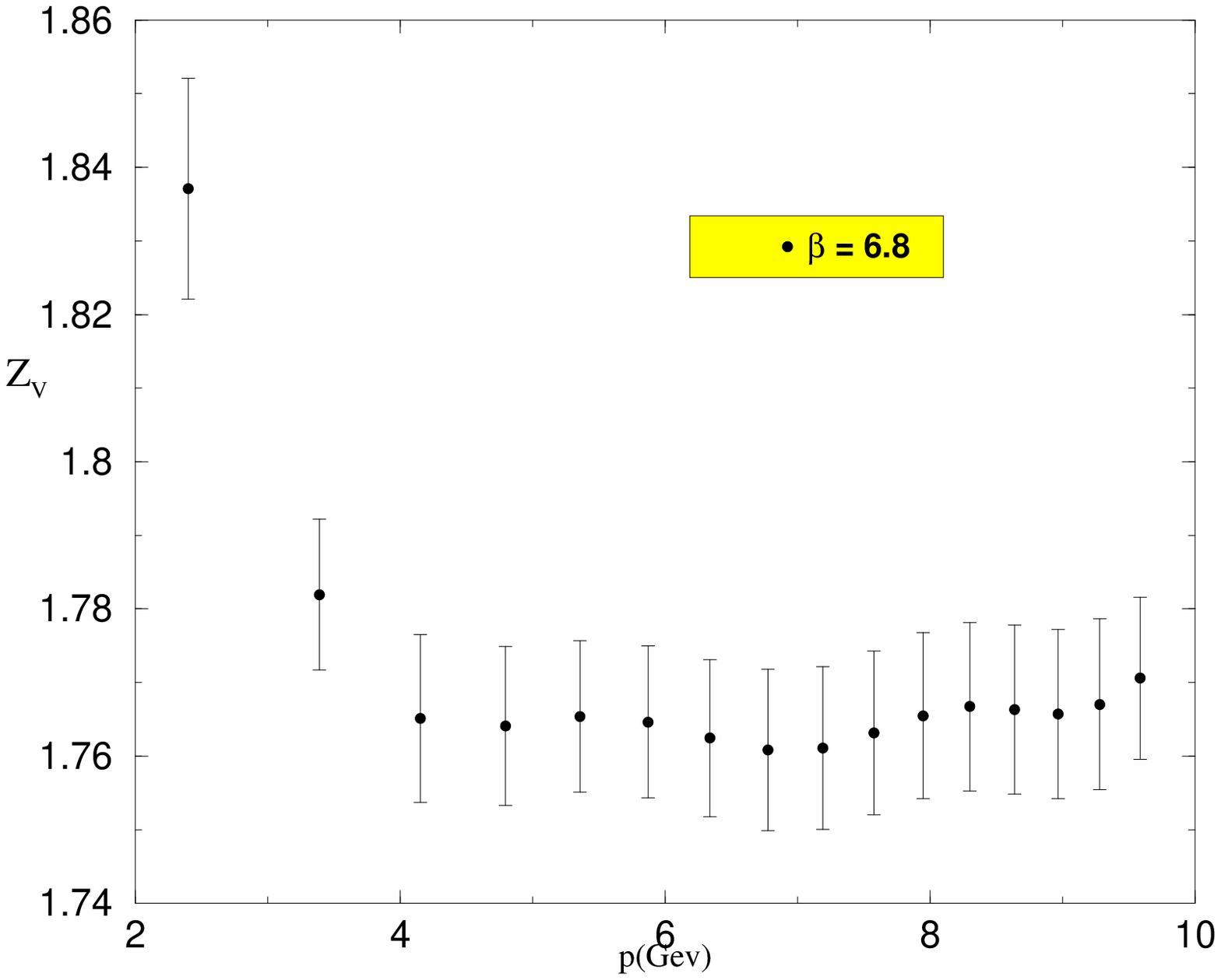,height=6cm}}
\caption{\small Lattice data for overlap-$Z_V$ 
after application of the `` $\mathbf p^{[2n]}$ 
extrapolation method '' for $\beta=6.0$ and 6.8.}
\label{Fig:Zv-over-R}
\end{center}
\end{figure} 
%%\vspace*{-4.5cm}
%%%%%%%%%%%%%%%%%%%%%%%%%%%%%%%%

%%%%%%%%%%%%%%%%%%%%%%%%%%%%%%%%%%%%%%%%%%%%%%%%%%%%%%%%%%%%%%%%%%%%%%
%%%%%%%Figure 7
%%%%%%%%%%%%%%%%%%%%%%%%%%%%%%%%%%%%%%%%%%%%%%%%%%%%%%%%%%%%%%%%%%%%%%
\begin{figure}[t]
\begin{center}
\leavevmode
\mbox{\epsfig{file=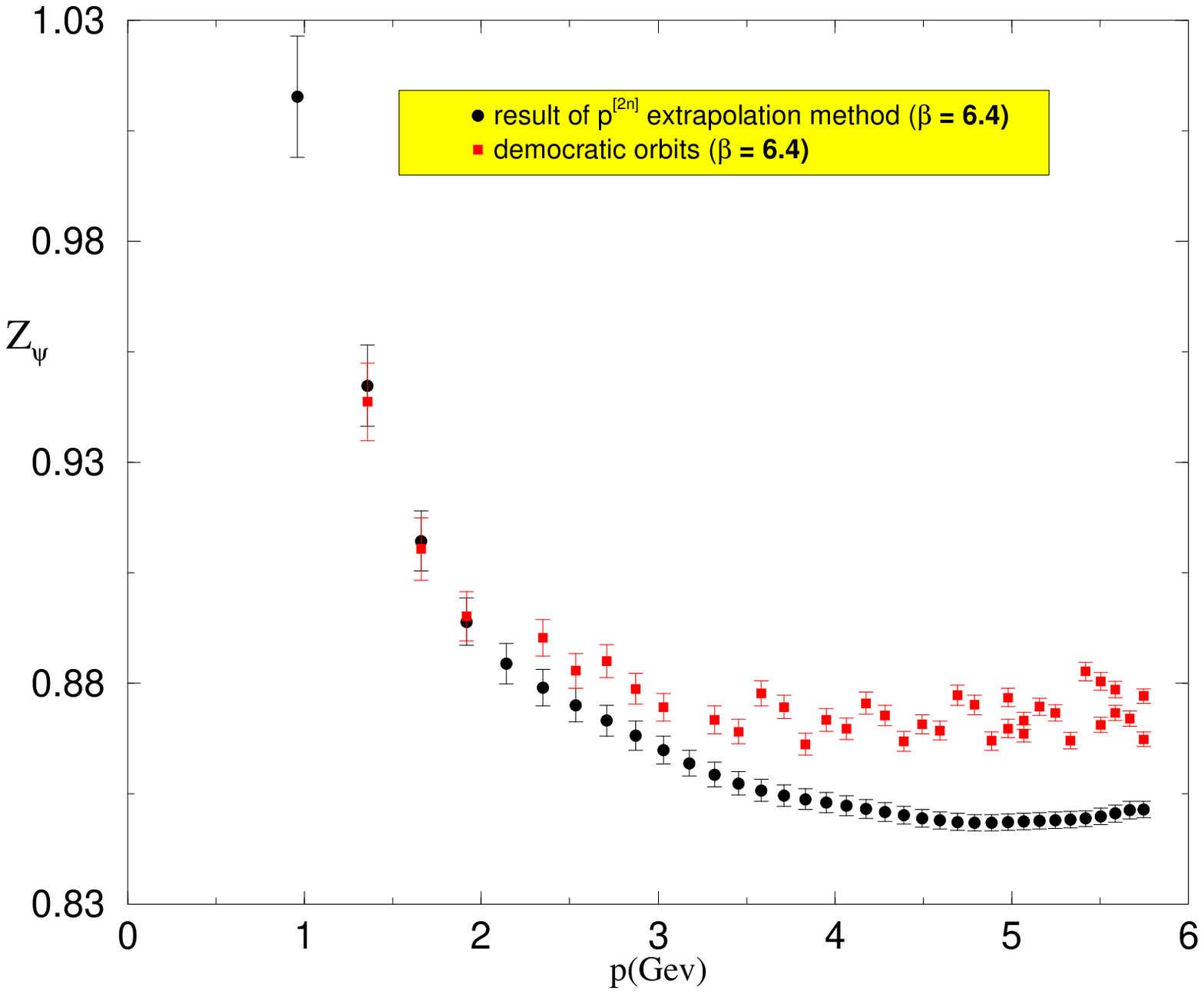,height=6cm}}
\mbox{\epsfig{file=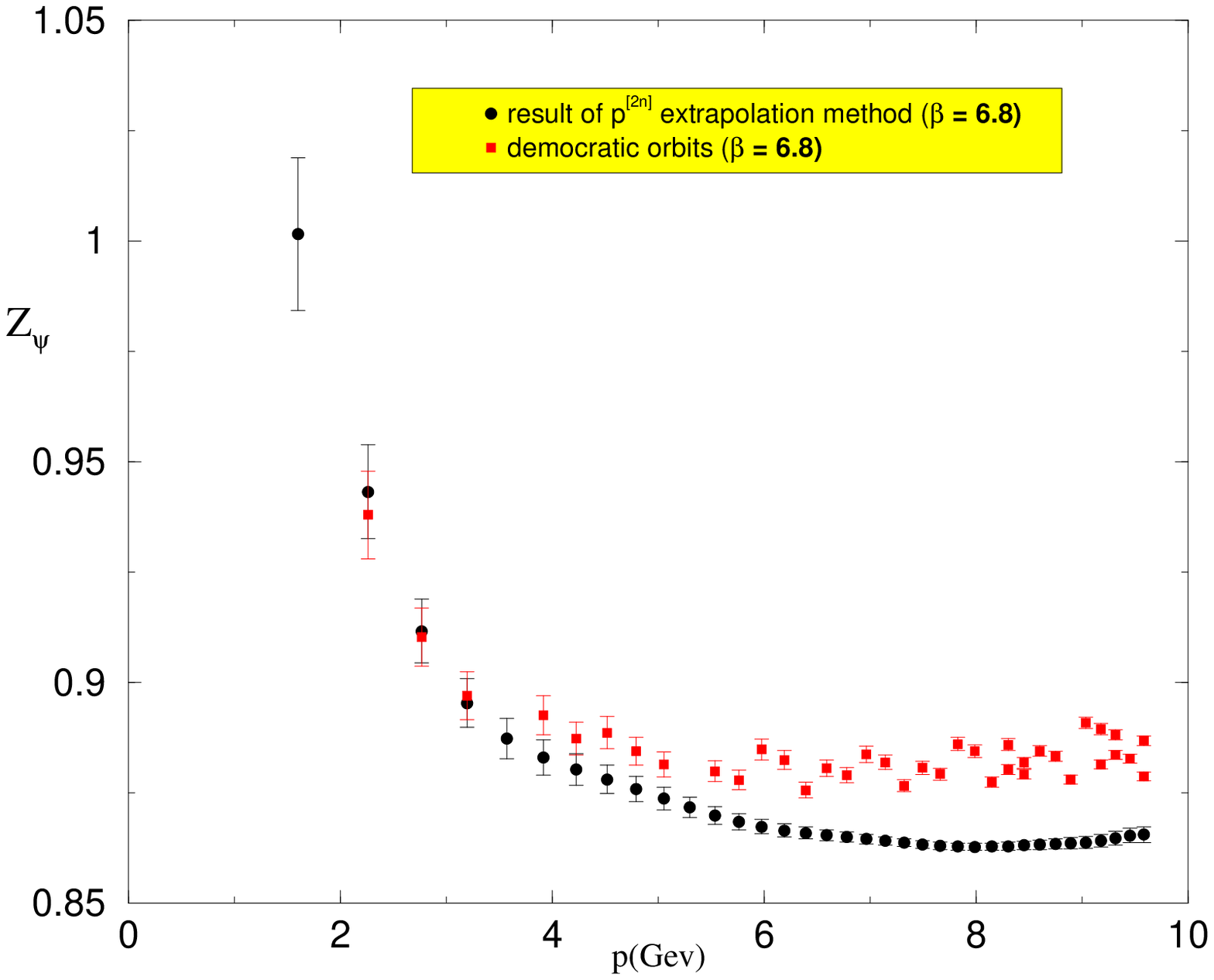,height=6cm}}
\caption{\small 
Red squares represent the ``democratic'' orbits, defined by
$p^{[4]}/(p^2)^2 \le .5$, for
clover-$Z_\psi$ at $\beta=6.4, 6.8$. 
The black circles represent the result of the $p^{[2n]}$
extrapolation method. The latter exhibit a much smoother
 behavior.}
\label{Figdem}
\end{center}
\end{figure} 
%%\vspace*{-4.5cm}
%%%%%%%%%%%%%%%%%%%%%%%%%%%%%%%%

%%%%%%%%%%%%%%%%%%%%%%%%%%%%%%%%%%%%%%%%%%%%%%%%%%%%%%%%%%%%%%%%%%%%%%
%%%%%%%Figure 8
%%%%%%%%%%%%%%%%%%%%%%%%%%%%%%%%%%%%%%%%%%%%%%%%%%%%%%%%%%%%%%%%%%%%%%
\vskip 2cm
\begin{figure}[hbt]
\begin{center}
\leavevmode
\mbox{\epsfig{file=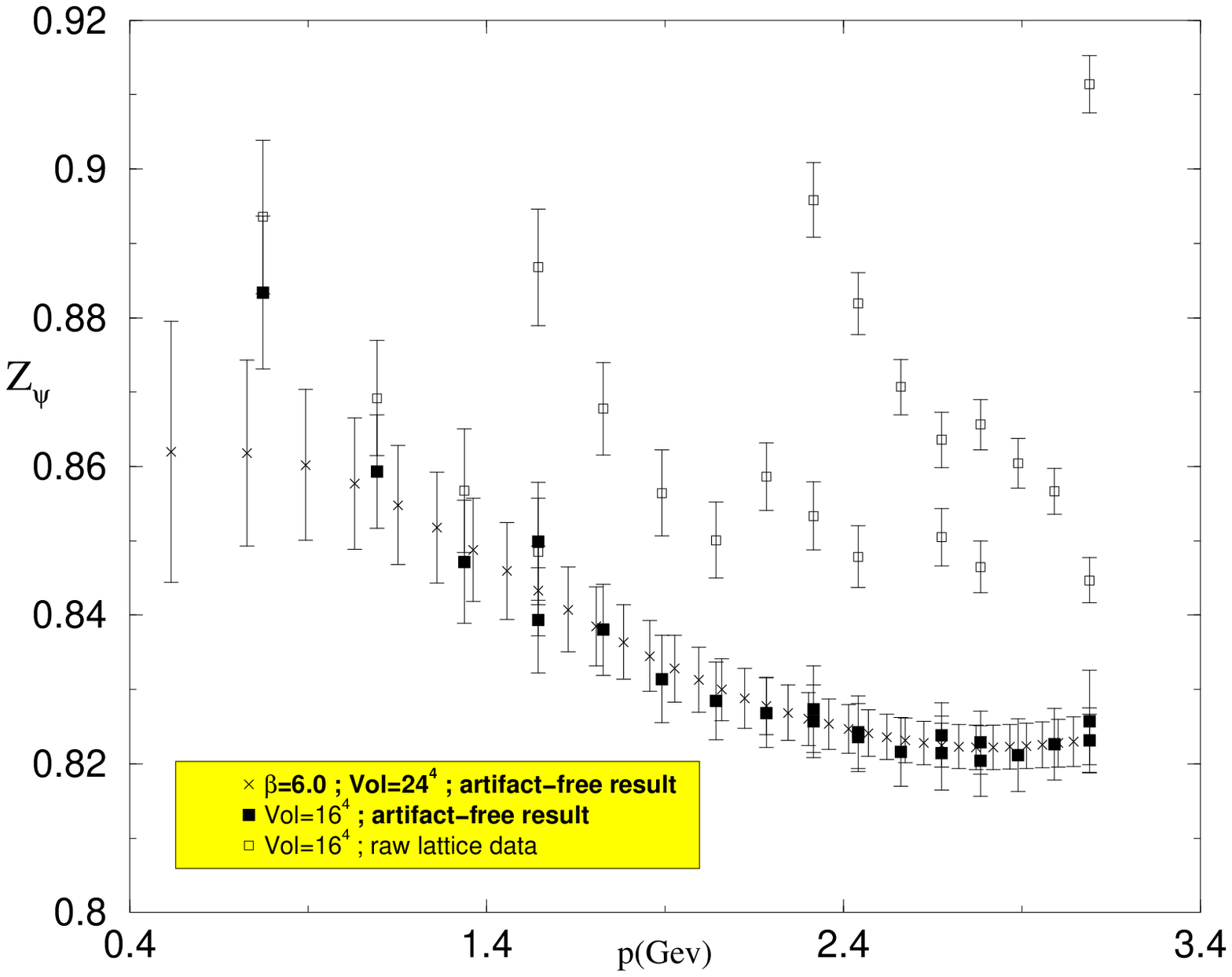,height=6cm}}
\mbox{\epsfig{file=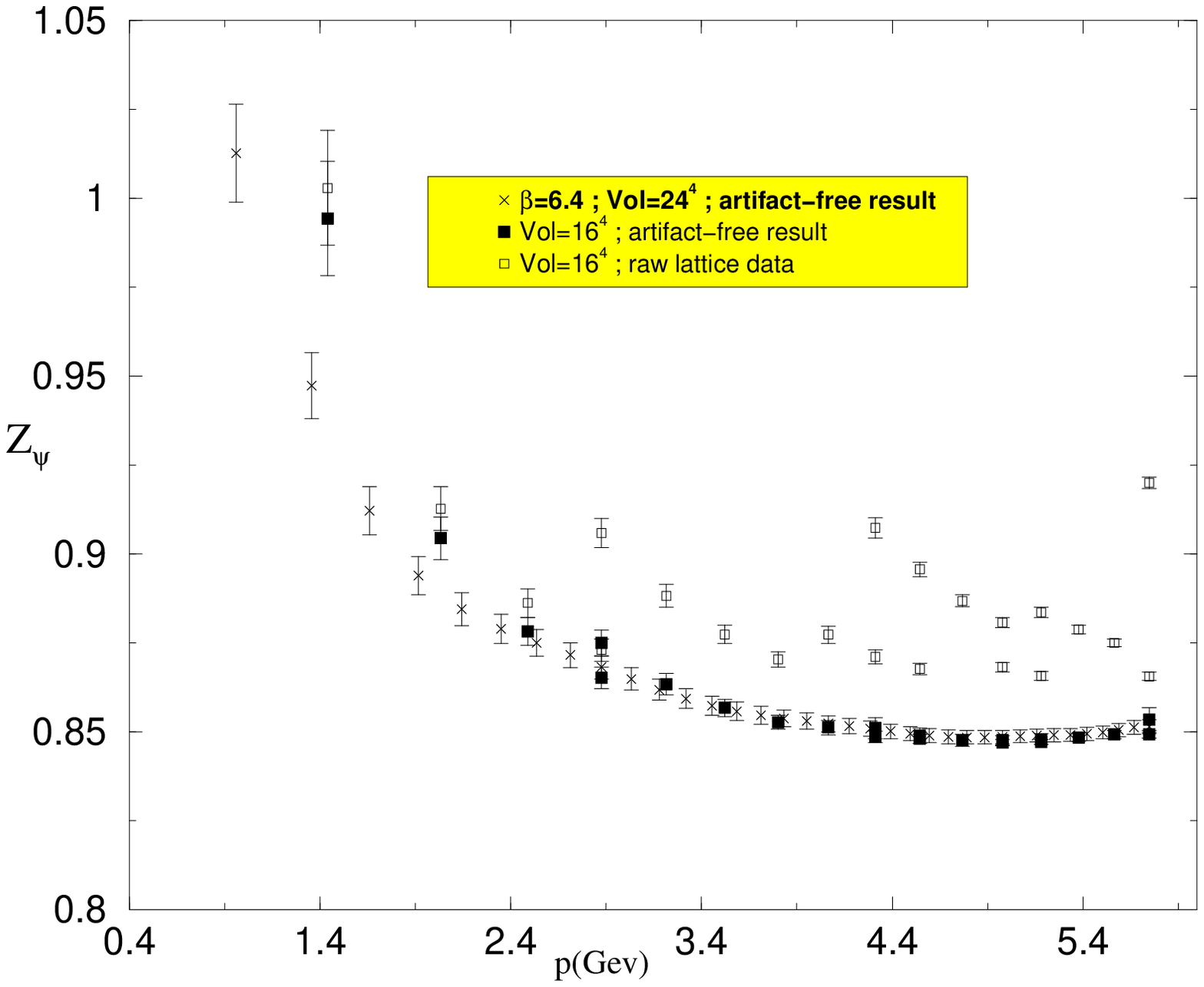,height=6cm}}
%\vskip -2.0 cm
\caption{\small The white squares show the raw lattice results
for clover-$Z_\psi$ with a volume of $16^4$. The black squares are the same
after the artifacts computed  with a volume of $24^4$ have been 
subtracted and the crosses represent the artifact-free result 
computed with a volume of $24^4$. The agreement between black 
squares and crosses is striking except for  the
smallest momentum on $16^4$. The figure to the left (right) is for
$\beta=6.0$  ($\beta=6.4$).}
\label{Fig16}
\end{center}
\end{figure} 
%%\vspace*{-4.5cm}
%%%%%%%%%%%%%%%%%%%%%%%%%%%%%%%%

%============================================
%============================================
\section{Lattice calculations}
\label{lattice}
We have used  improved Wilson quarks (often called clover) with the
CSW coefficients computed in~\cite{Luscher:1996ug}. 100 gauge 
configurations have been computed at $\beta=6.0, 6.4, 6.6,6.8$ with
volumes $24^4$, $16^4$ and $8^4$. We have performed the  calculation
for five quark masses but in practice, for what is our concern in this
paper, the quark mass  dependence has non surprisingly proven to be
negligible and  for simplicity we will only present the results for
the lightest quark mass, about 50 MeV, {\it i.e.}
\bea
 \kappa= 0.1346,\,\,  0.13538,\,\,
0.13515,\,\,  0.13489 \quad {\rm for} \quad
\beta = 6.0,\,\, 6.4,\,\,6.6,\,\,6.8 
\eea 
 It should also be mentioned  
that all the results presented refer to the $24^4$ lattices unless 
 stated otherwise. 

We have also used overlap 
fermions~\cite{Neuberger:1997fp,Capitani:1999uz} with 
about the same  mass {\it i.e.}
\bea a\,m_0= 0.03,\,\,0.01667,\,\,0.01.\,\quad {\rm for} \quad
\beta = 6.0,\,\, 6.4,\,\,6.8 
\eea
with $s=0$ and volumes of only $16^4$ due to memory limitations.  
 The bare mass $m_0$ and $s$ are defined from
\bea\label{overlap}
D_{\rm over}=  ( 1+s + am_0/2) + ( 1+s - am_0/2) 
\frac{D_{\rm w}(-(1+s))}
{\sqrt{D_{\rm w}(-(1+s))^\dagger \, D_{\rm w}(-(1+s))}}
\eea

where $ D_{\rm w}(-(1+s))$ is the Wilson-Dirac operator with a (negative)
mass term $-1-s$
\bea
  D_{\rm w}(-1-s) \equiv \frac{1}{2} \gamma_\mu ( \nabla_\mu +\nabla^*_\mu)
  - \frac{1}{2} a \nabla^*_\mu \nabla_\mu -1-s \,,
\eea

 The propagators $S(x,0)$ from the origin to point $x$ have 
 been computed and their Fourier transform
\bea
\widetilde S(p) = \sum_{x} e^{-i p\cdot x} S(x,0)
\eea
have been averaged among all configurations and all momenta $p_\mu$ within
one orbit of the hypercubic symmetry group of the lattice, 
exactly as  for gluon Green functions  
in~\cite{Boucaud:1998bq}-\cite{Becirevic:1999hj}. 
In the case of overlap quarks 
the propagator is improved according to a standard 
procedure~\cite{Capitani:1999uz} 
which eliminates $O(a)$ discretization errors:
\bea\label{impr}
\widetilde S_\ast(p) = \frac {\widetilde S(p) - \frac 1 2}{1-am_0/2}
\eea
From now on, the notation $S(p)$ will represent the improved 
quark propagator in the case of overlap  quarks and the standard
 one in the case of clover quarks.

In both cases we fit the inverse quark propagator by
\bea
\widetilde S^{-1}(p) = \delta_{a,b}\, {Z_\psi(p^2)}\,\left(
 i\,\bar \pslash + m(p^2)\right)
\eea
according to eq.~(\ref{Sm1}) and where $\bar p_\mu$ is defined in 
eq. (\ref{tilbar}). The three point Green functions
with vanishing momentum transfer are computed by averaging analogously over
the thermalised configurations and the points in each orbit
\bea
G_\mu(p, q=0)= <\gamma_5 \,\widetilde S(p)^\dagger \,\gamma_5 \,
\gamma_\mu \,\widetilde S(p)>.
\eea
where the identity $S(0,x)= \gamma_5 S^
\dagger(x,0) \gamma_5$ has been used.
The vertex function is then computed according to eq.~(\ref{gammamu1}) and 
we choose for the lattice form factor  $g_1$ :
\bea
g_1(p^2) = \frac 1 {36}{\rm Tr} \left[ \Gamma_\mu(p, q=0) 
\left(\gamma_\mu - \bar p_\mu \frac{\bar \pslash}{\bar p^2} 
\right)\right]
\eea
where the trace is understood over both color and Dirac indices.

Finally, according to the Ward identity (\ref{ward2}) we compute 
$Z_V$ simply from
\bea\label{defzv}
Z_V(p^2) \equiv \frac {Z_\psi(p^2)}{g_1(p^2)}.
\eea
where the $p^2$-dependence of $Z_V$ coming from lattice artifacts 
has been explicitly written.

%We have already stressed that $Z_V$ should not depend on $p^2$. 
%However, the lattice artifacts can generate a
%$p^2$ dependence of the ratio $Z_\psi(p^2)/g_1(p^2)$ at the level 
%of the raw lattice data. This is why we write such a dependence in
%eq. (\ref{defzv}). 

In whole this paper we will use the values in the following table
\ref{spacing} for the lattice
spacings, which follow the $\beta$ dependence found in
ref.~\cite{Capitani:1998mq},
\begin{table}[h]
\begin{center}
\begin{tabular}{||c||c|c|c|c||}
\hline
\hline
$\beta$ & 6.0 & 6.4 & 6.6 & 6.8 \\
\hline
$a^{-1}$ (GeV) & 1.966 & 3.66 & 4.744 & 6.1 \\
\hline
$a$ (fm) & 0.101 & 0.055 & 0.042 & 0.033 \\
\hline
\hline
\end{tabular}
\end{center}
\caption{{\small Lattices spacings}}\label{spacing}
\end{table}

%\bea\label{spacing}
%a^{-1}(6.0)&=&1.966 \,{\rm GeV},\quad a^{-1}(6.4)=3.66 \,{\rm GeV},\quad
%a^{-1}(6.6)= 4.744\, {\rm GeV},\quad
%a^{-1}(6.8)=6.1 {\rm GeV}, \nonumber   \\
%a(6.0)&=&0.101 \,{\rm fm},\quad a(6.4)=0.055 \,{\rm fm},\quad
%a(6.6)=0.042 \,{\rm fm},\quad
%a(6.8)=0.033 \,{\rm fm}.
%\eea

%============================================
%============================================
\section{Elimination of lattice hypercubic artifacts}
\label{arti}

The question of eliminating lattice artifacts 
has been our main difficulty in the accurate study of 
the quark propagator. We became convinced that it was 
absolutely impossible to say anything sensible without
an extremely careful elimination of artifacts. Here we
mean mainly the ultraviolet artifacts, the 
infrared artifacts having never been really troublesome
in this problem. 

We have elaborated a very powerful method to deal with
hypercubic artifacts i.e. with those ultraviolet artifacts 
which come from the difference between the hypercubic 
geometry of the lattice and the fully hyper-spherically 
symmetric  one  of the continuum Euclidean space. The 
principle of this method is based on identifying the 
artifacts which are invariant for the $H_4$ symmetry of 
the hypercube, but not for the $SO(4)$ symmetry of the
continuum.  

Once these artifacts have been eliminated it is obvious, as we shall
show,  that other - $SO(4)$-invariant ultraviolet artifacts - are
present.  And these turn out to be even trickier to deal with,
mainly because we did not fully understand their rationale. 

Therefore, we intend to restrict ourselves in this  paper to a
careful explanation of the hypercubic artifacts  elimination method
since we believe it represents a real progress and it can be useful 
for many other lattice calculations.  The treatment of the 
$SO(4)$-invariant artifacts and of the physical results concerning the 
quark propagator will be given in a later publication. 

%============================================
\subsection{$\mathbf p^{[2n]}$ extrapolation method}

Since we use hypercubic lattices our results are invariant for 
a discrete symmetry group, $H_4$, a  subgroup of the 
continuum Euclidean $SO(4)$. This implies  that lattice data
for momenta which are not related by  an $H_4$ transformation
but are by a $SO(4)$ rotation will in principle differ.
Of course this difference must vanish 
when $a\to 0$ but it must be considered among the 
discretisation effects, i.e. ultraviolet artifacts. 
For example, in perturbative lattice calculations one encounters 
the expressions
\bea\label{tilbar}
\tilde p_\mu \equiv \frac 2 a \sin \left(\frac {a p_\mu} 2\right),\quad
\bar p_\mu \equiv \frac 1 a \sin \left( a p_\mu \right).
\eea
Both are equal to $p_\mu$ up to lattice artifacts: 
\bea\label{tilbar4}
\tilde {p^2} \equiv \sum_{\mu=1,4}\tilde p_\mu^2 = p^2 - \frac 1 {12} a^2 p^{[4]}
+ \cdots  \quad \bar p^2 =   p^2 - \frac 1 {3} a^2 p^{[4]}+ \cdots\,,\, {\rm
where}\, p^{[2n]} \equiv \sum_{\mu=1,4} p_\mu^{2n}\, .
\eea
All terms in the dots are proportional to  $a^{2n} p^{[2n+2]}$ .
$p^2$, $\tilde p^2$, $\bar p^2$, $a^2 p^{[4]}$ are invariant under $H_4$
but only $p^2$ is under $SO(4)$. For example the momenta $2\pi (1,1,1,1)\,/L$
and $ 2\pi (2,0,0,0)\,/L$ have the same $p^2$ but different $p^{[4]}, \tilde p^2$
and $\bar p^2$. In other words, if we call an orbit the set of momenta related by  $H_4$
transformations, different orbits, corresponding to the same $p^2$, will 
in general have different $p^{[4]}$.
The hypercubic artifacts can be detected by looking carefully
for a given quantity at a given $p^2$ how it depends on the orbit.
 
One method proposed with success for the gluon 
propagator~\cite{Becirevic:1999uc}, \cite{Becirevic:1999hj}
analyses a generic lattice measured quantity $Q$ as a function $Q(p^2, 
p^{[4]})$. For a given value of $p^2$, if enough  different values of $p^{[4]}$
exist the quantity $Q$ is fitted by 
\bea\label{roiesnel}
Q(p^2, p^{[4]} ) = Q(p^2, 0) + \frac{\partial Q(p^2, y)}{\partial y}|_{y=0} 
 \,\,{p^{[4]}}.
\eea
where $Q(p^2, 0)$ is free of hypercubic artifacts and where
$\frac{\partial Q(p^2, y)}{\partial y}|_{y=0}$ is computed numerically
for each $p^2$ from the slope of the lattice data for $Q(p^2,  p^{[4]})$ 
as a function of $p^{[4]}$. Of course, we could also consider $p^{[6]}$,
etc.  but usually there are not enough different orbits for one
given $p^2$ to fit more than the $p^{[4]}$ correction.

Let us call this method the {\bf ``$\mathbf p^{[4]}$
extrapolation method''}. For the gluon propagator this method has been 
shown~\cite{Becirevic:1999hj} to lead to a resulting function
 $G(p^2,  p^{[4]}=0)$  much smoother than  the direct lattice
results $Q(p^2, p^{[4]})$, even if the latter are restricted, as often done,
to the ``democratic'' momenta, i.e. to those which have the smallest $p^{[4]}$. 
This method could be applied with some success to the function clover-$g_1$ 
(i.e. $g_1$ obtained from improved Wilson quarks). 

But in general, as we shall see, when applied to clover-$Z_\psi$ or to 
the quantities computed from overlap quarks, the { ``$\mathbf p^{[4]}$
extrapolation''} method fails. The signal
of this failure is that the resulting function  $Q(p^2,  p^{[4]}=0)$
still shows sizable  oscillations typical of hypercubic artifacts. 

We then propose a {\bf  ``$\mathbf {p^{[2n]}}$
extrapolation method''} which allows to eliminate much more efficiently 
these hypercubic artifacts. The improvement goes in two directions:

i) Instead of fitting the $p^{[4]}$ slope separately for each 
value of $p^2$ we try a global fit of the hypercubic artifacts over all values
of $p^2$  

ii) We chase hypercubic artifacts up to order $a^4$. 

In order to perform a global fit we start from the remark that in
 this paper we are dealing with dimensionless quantities,
$g_1$ and $Z_\psi$. 
It is thus natural to expect that  hypercubic 
artifacts contribute via dimensionless quantities times a 
constant~\footnote{We neglect a possible logarithmic dependence in $p^2$.}.
Next we assume that there is a regular continuum
limit which implies that in the denominator we can have only 
physical quantities, namely~\footnote{ In all this discussion we consider
the mass as negligible.} a function of $p^2$.  These two priors lead us
 to a Taylor expansion with terms of the type  
\bea\label{ketn}
\left(\frac{a^{2k} p^{[2k+2n]}}{(p^2)^n}\right)^m \qquad 
k>0,\quad n\ge 0\quad k+n >1,\quad m>0.
\eea

This still leaves us with far too many terms to make sensible fits. 
It is reasonable to truncate this series in $a$ and we choose to 
expand it up to $a^4$.
 We will also truncate it to $n \le 1$, and now comes
 an heuristic argument to justify this truncation. 
 
The lattice results are $H_4$ invariant and thus typically functions of
$\tilde p^2$, $\bar p^2$ and $\tilde p\cdot \bar p$.  Dimensionless quantities 
will depend on ratios of these quantities and on terms~\footnote{ Terms like  
 $a^4 (\tilde p\cdot \bar p) \tilde p^2$ yield only ${\cal O}(a^6)$
 anisotropic terms.} like $a^2 \tilde
p^2$. An examination of these
 shows that they produce  anisotropic terms 
corresponding to $n=0$ and $n=1$ in the classification of eq. (\ref{ketn}). 
For example, $a^2 \tilde p^2$ produces from the expansion
 eq. (\ref{tilbar4}) a term $\propto a^4 p^{[4]}$
while $\tilde p\cdot \bar p/ \bar p^2$ produces terms $\propto a^2 p^{[4]}/p^2$
etc. This is also what is found in the free case~\cite{Capitani:1999uz}.

Hence we fit the data according to
\bea\label{micheli2-1}
Q(p^2, a^2 p^{[4]},a^4 p^{[6]},\cdots) = Q(p^2, 0, 0) + 
c_1 \, \frac{a^2 p^{[4]}}{p^2}
+ c_2\,
\left(\frac{a^2 p^{[4]}}{p^2}\right)^2 + c_3\, \frac{a^4 p^{[6]}}{p^2} 
+ c_4\, a^4  p^{[4]}.
\eea

\begin{figure}[h]
\begin{center}
\leavevmode
\mbox{\epsfig{file=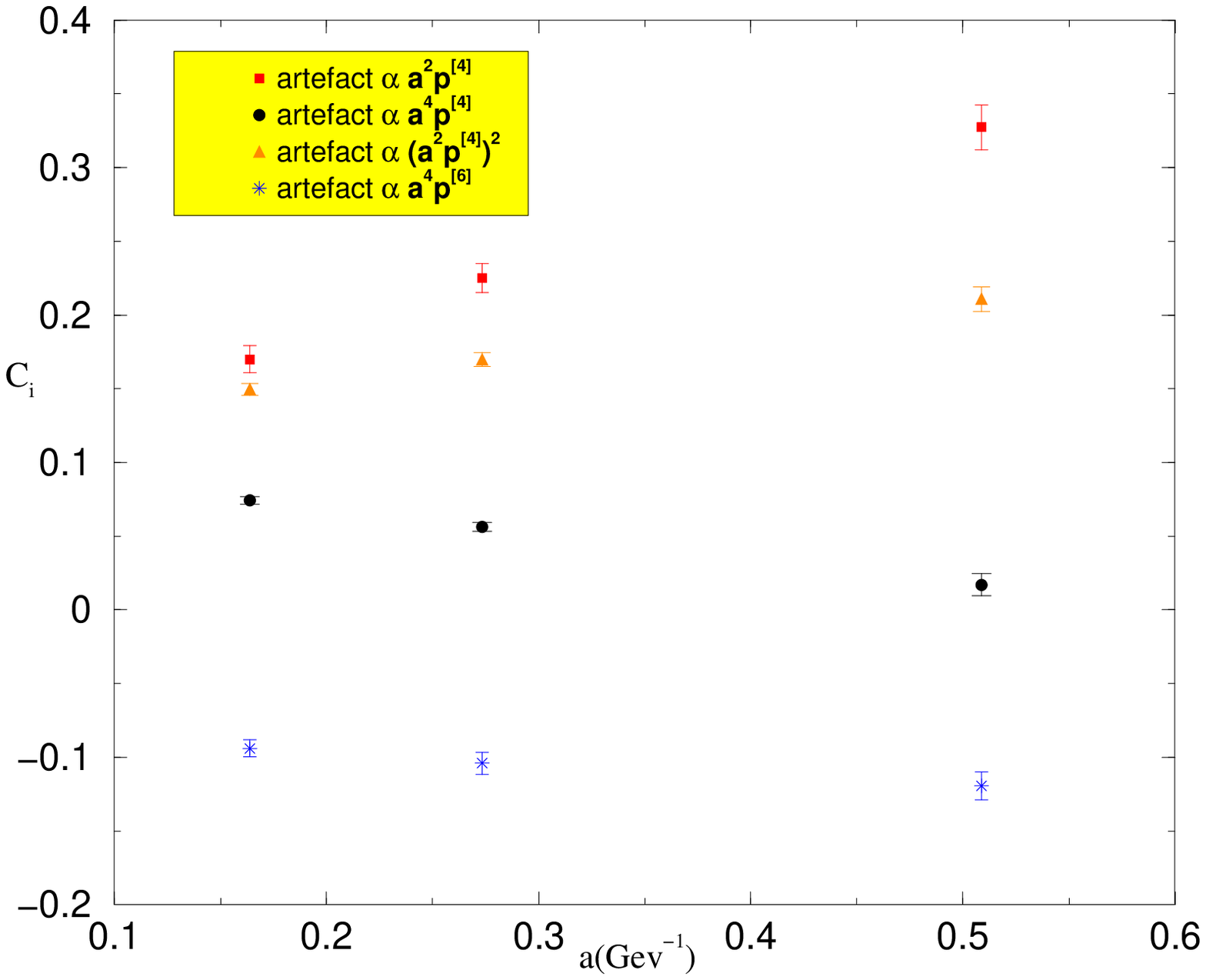,height=6cm}}
\mbox{\epsfig{file=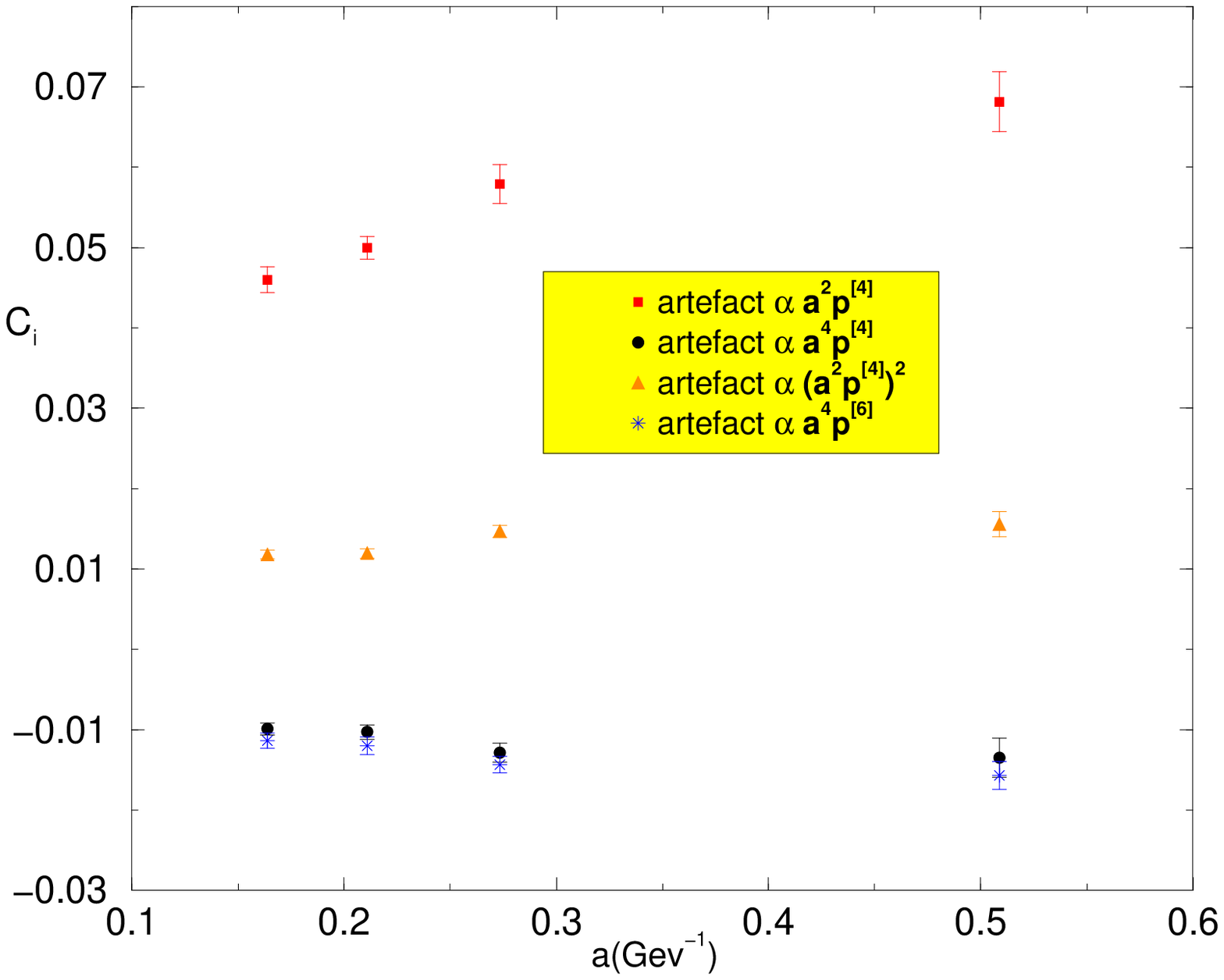,height=6cm}}
%\vskip -2.0 cm
\caption{\small Coefficients of the hypercubic
artifacts for $Z_\psi$ from overlap quarks (left) and clover (right)
 as a function
of $a$ for $\beta=6.0,6.4,6.8$ and 6.6 in the clover case. The  
squares  corresponds to $c_1$, the coefficient
of $a^2p^{[4]}/p^2$, the triangles to $c_2$, the coefficient
of $(a^2p^{[4]}/p^2)^2$, the stars to $c_3$, the coefficient
of $(a^4p^{[6]}/p^2)^2$ and the circles to $c_4$, the coefficient of 
$a^4p^{[4]}$. It suggests a linear dependence
on a, especially for the overlap quarks. The overlap artifacts
are larger than the clover ones by one order of magnitude.
}
\label{coefs}
\end{center}
\end{figure}

A remark is needed about the $a$ dependence of the coefficients 
$c_i$. Being dimensionless it is expected in perturbation
theory that these coefficients depend only logarithmically on $a$
and taking them as constants would seem reasonable. This conjecture does
not work as shown in fig. \ref{coefs} which is the sign of non-perturbative 
${\cal O}(a \Lambda_{\rm QCD})$ contributions. Still this figure shows a rather 
convincing linear dependence of the $c_i$'s for $Z_\psi$  which tells that
 a good global fit 
can be performed by expanding the coefficients: $c_i(a) = c_i^0 + a c_i^1$.
For $g_1$ the $a$ dependence is not linear while the hypercubic 
artifacts are one order of magnitude smaller.  

%%%%%%%%%%%%%%%%%%%%%%%%%%%%%%%%%%%%%%%%%%%%%%%%%%%%%%%%%%%%%%%%%%%%%%
%%%%%%%Figure 9
%%%%%%%%%%%%%%%%%%%%%%%%%%%%%%%%%%%%%%%%%%%%%%%%%%%%%%%%%%%%%%%%%%%%%%
\vskip 1cm
%%\vspace*{-4.5cm}
%%%%%%%%%%%%%%%%%%%%%%%%%%%%%%%%

The functional form used for $Q(p^2, 0, 0)$ does not influence significantly 
the resulting artifact coefficients. We can even avoid using any assumption 
about this functional form by taking all the values for $Q(p^2, 0, 0)$
as parameters which can be fitted~\footnote{We have enough data for that.}.

This improved correction of hypercubic artifacts turned out to be particularly 
necessary for $Z_\psi$. 
 In fig.~\ref{FigH4} the very strong hypercubic artifacts produce an impressive 
branched structure with a kind of periodicity.
In fig. \ref{FigRM}  we show the effect of both the use of eqs.
(\ref{roiesnel}) and (\ref{micheli2-1}). It  
turns out to that {\bf ``$\mathbf p^{[4]}$ extrapolation method''},
eq. (\ref{roiesnel}),  makes the branches disappear and the curve
 look much smoother. However it still contains some oscillations reminiscent
of the hypercubic artifacts. The {\bf ``$\mathbf p^{[2n]}$ 
extrapolation method''}, eq. (\ref{micheli2-1}), brings in a further dramatic
smoothing.   

The same is true for overlap-computed quantities. In figs. \ref{Fig:Zpsi-over} and 
 \ref{Fig:Zv-over} the raw lattice data for $Z_\psi$ and $Z_V$ exhibit
 dramatically the ``half-fishbone'' structure which is a symptom
  of strong hypercubic artifacts. In figs. \ref{Fig:Zpsi-over-R} and 
 \ref{Fig:Zv-over-R} the same data are shown after applying the $\mathbf p^{[2n]}$ 
extrapolation method. Clearly the curves are now perfectly smooth. 
We will return later to the fact that $Z_V$ is not a constant. 
   
Altogether we would like to stress the following hierarchy:
First, the hypercubic artifacts are one order of magnitude larger for 
overlap quarks  than for clover ones, see fig.~\ref{coefs}. Second, 
for both types of quarks the hypercubic artifacts
for $Z_\psi$ are one order of magnitude larger than those for 
$g_1$.

%============================================
\subsection{Comparison with the ``democratic''  method}

The hypercubic artifacts, sometimes called ``anisotropy artifacts'' have been a
long standing problem in lattice calculations.  Studying the gluon propagator 
the authors of ref. \cite{Leinweber:1998im} where aware that the problem was
related to the fact that for a given momentum $p^2$ these artifacts were
minimised when the components were as small as possible, i.e. such that the
components are not too hierarchical, and that the ideal situation was the
diagonal $p\propto (1,1,1,1)$, whence the name commonly used of a ``democratic''
repartition of the momentum in all directions. Therefore they have proposed  a
selection   keeping only the orbits having a point within a cylinder around the
diagonal. Several other criteria have been used.

In this subsection we want to compare  this method of eliminating the 
non-democratic points to the  `` $\mathbf a^2 p^{[2n]}$
extrapolation method'', eq.  (\ref{micheli2-1}). 
If we try  to select, \cite{Leinweber:1998im}, the orbits which are in
a cylinder around the diagonal with a radius $2\pi/L$, we are left with
only 11 orbits among 69. 

In  order to  have a less restrictive criterion and to make the bridge with
the method used here we  will use the $p^{[2n]}$'s defined in eq. (\ref{tilbar4}).
In our language, democracy can be translated  as a small  enough ratio
$p^{[4]}/(p^2)^2$. 
Momenta proportional to $(1,1,1,1)$ and $(1,0,0,0)$ have ratios $1/4$ (minimum) 
and $1$ (maximum) respectively.
%$p\propto (1,1,1,1)$ ($p\propto (1,0,0,0)$) corresponds to
%$p^{[4]}/(p^2)^2 = 1/4$ ($p^{[4]}/(p^2)^2 = 1$)  which is  the minimum
%(maximum).  
In fig. \ref{Figdem} we plot for $Z_\psi$ the result 
of the following fit. We  take the ``democratic'' orbits defined  by
$p^{[4]}/(p^2)^2 \le .5$. This leaves 40 orbits out of 69 for every $\beta$.
Fig. \ref{Figdem} clearly  shows oscillations demonstrating
that the hypercubic artifacts have not been totally eliminated. 
 For this reason and also because of the
loss of information due to the rejection of ``undemocratic'' points, we 
did not use this method.

%============================================
\subsection{Finite volume artifacts}

We did not see any sizable finite volume effect {\it in the case of clover
quarks}. To illustrate 
this claim we have performed the following exercise illustrated in fig.
\ref{Fig16}. We have subtracted from the raw lattice results 
  clover-$Z_\psi$, computed with a volume of $16^4$,  the artifacts  
with the coefficients $c_1,..,c_4$ fitted on a volume $24^4$, namely 
the results of eq.~(\ref{micheli2-1}) and compared the result to
the artifact-free function  $Q(p^2, 0, 0)$ computed with $24^4$. 
The agreement as shown in  fig. \ref{Fig16} is impressive except for  the
smallest momentum on $16^4$.
We have also checked on several examples that the inclusion in the fits
of finite volume artifacts of the type $1/(L^2p^2)$ 
did not produce any significant change in the results.

%============================================
\subsection{$Z_V$ and other discretization artifacts}
\label{other}

Figure \ref{Fig:Zv-over-R} presents the result of extracting the
hypercubic artifacts from the raw lattice data for $Z_V$ computed 
according to eq. (\ref{defzv}). It shows  that we are not through with
 artifacts. Indeed,  as we have already mentioned, the
artifact-free $Z_V$ must not depend  on $p^2$. It is expected to depend
on the bare coupling constant  i.e. on $\beta$ but not on the momentum.
The figures \ref{Fig:Zv-over-R} show smooth curves which confirms the
efficient elimination of  hypercubic artifacts, but it also shows a
residual significant dependence on $p^2$ up to 50 \% variation in the
case of overlap quarks at $\beta = 6.0$. This dependence is necessarily
due, either to additional finite lattice spacing artifacts which are not
of the hypercubic type but are $SO(4)$-invariant, or to finite volume
artifacts. It is however noticeable that at $\beta=6.8$ $Z_V$ is 
really flat except for the two first points.  

These artifacts do not look simple since $Z_V$ increases 
 at small $p^2$.  We have discarded finite
volume effects in the preceding section for clover quarks but we could
not, by lack of computing resources, perform the same check  for overlap
quarks. In particular the strong $p^2$ dependence of $Z_V$ at small
momentum seen in fig. \ref{Fig:Zv-over-R} is evocative of $\propto
1/(L^2p^2)$  finite volume effects.   But these type of effects 
would produce exactly the same shape at both $\beta$'s. 
The difference between 
the two plots in fig. \ref{Fig:Zv-over-R} shows that the
case  is more subtle  and
small momentum ultraviolet artifacts can also be present. This clearly
needs a careful study and some theoretical understanding which will be
developed elsewhere.

Finally it is useful to notice that the $p^2$ dependence of $Z_V$
is one order of magnitude larger for overlap quarks than 
for clover ones.

%============================================
%============================================
\section{Discussion and conclusions}
\label{concl}

We have computed the quark field renormalisation constant $Z_\psi$ and
the vector current form factor $g_1(0)$  both with improved Wilson
quarks (clover) and with overlap quarks. The quark
propagator is very strongly  affected by lattice artifacts~\footnote
{Notice that the size of lattice artifacts might depend strongly
on the parameter $s$ defined in eq.(\ref{overlap})
and which we have taken to vanish in this paper.}. This is
already very annoying in the case  of clover quarks but is even
one order of magnitude larger for overlap quarks. See for example
figures~\ref{FigH4}  and~\ref{Fig:Zpsi-over}. 
Hypercubic artifacts mainly affect $Z_\psi$, while the vertex
factor $g_1$ is less affected by one order of magnitude. 
This forwards an invitation to use intensively Ward identities 
and vertex functions simultaneously to the propagator. 

In order to eliminate the latter hypercubic  artifacts we have
improved the method presented in
refs.~\cite{Boucaud:1998bq},~\cite{Becirevic:1999uc} into what we call
the  ``$\mathbf {p^{[2n]}}$ extrapolation method''. It is based on a
systematic expansion over the invariants of the hypercubic group $H_4$
which are not invariants of the $SO(4)$ symmetry group of the 
Euclidean continuum and on a systematic use of dimensional arguments
to guess the $p^2$ dependence of the artifacts.  We have shown that
 this method totally 
eliminates the dramatic disorder of the raw data, see
figures~\ref{FigH4}-\ref{Fig:Zv-over}, which exhibit a shape vaguely
reminiscent of the half of a fishbone.  

After applying the ``$\mathbf {p^{[2n]}}$ extrapolation method''
we get results which are perfectly smooth,
figs.~\ref{FigRM}-\ref{Fig:Zv-over-R}. In particular the artifact
fig.~\ref{Fig:Zpsi-over-R} for overlap quarks should be compared
to the raw lattice data fig.~\ref{Fig:Zpsi-over}. In
fig.~\ref{FigH4} there is also a very good agreement between the
clover raw data (black circles) and the red squares  computed
from eq.~(\ref{micheli2-1}). We have shown that this method is
much more efficient than the  popular ``democratic'' one, and it
is also more systematic and  allows the use of {\it all the
lattice data} which improves the  statistics.  We have decided to
center this paper on this method because, although the point
might look technical, we believe that  it can be of great help
for the lattice community when  artifacts are large.

The next task is to find out an equally efficient method to eliminate
the isotropic artifacts. Their presence is obvious from
fig.~\ref{Fig:Zv-over-R} which shows a strong $p^2$ dependence of
$Z_V$ after the anisotropic artifacts have been eliminated by the
$\mathbf {p^{[2n]}}$ extrapolation  while $Z_V$ should be a
constant~\footnote{See the discussion at section \ref{theor}}. 
Contrarily to $Z_V$, $Z_\psi = Z_V\,g_1$ is expected to depend on
$p^2$ as an effect of the perturbative QCD running  and of the
nonperturbative $<A^2>$ condensate.  Interesting physics can be
learned from this dependence provided we manage to fully control the
isotropic artifacts.  The constancy of $Z_V$, being a strong
constraint, will be a significant check that this control has been
achieved.

\section*{Acknowledgement}

We are grateful to  Claude Roiesnel for being at the origin of the
hypercubic artifacts elimination method which is extended here. 
We also thank Michele Pepe for a participation at an early stage
of this work. This work was
supported in part by the European Network  "Hadron Phenomenology
from Lattice QCD'', HPRN-CT-2000-00145, by the spanish CICYT contract
FB1998-1111 and by Picasso agreement HF2000-0056. We have used for
this work  the APE1000
located  in the Centre de Ressources Informatiques (Paris-Sud,
Orsay) and purchased thanks to a funding from the Minist\`ere de
l'Education Nationale and the CNRS.

%============================================
%============================================

%%%%%%%%%%%%%%%%%%%%%%%%%%%%%%%%%%%%%%%%%%%%%%%%%%%%%%
 
%%%%%%%%%%%%%%%%%%%%%%%%%%%%%%%%%%%%%%%%%%%%%%%%%%%%%%%%%


\vspace*{1.7cm}
{\small 
\begin{thebibliography}{99}
\bibitem{lane} K. D. Lane, {\it Phys. Rev.} {\bf D 10} (1974) 2605;
H. Pagels, {\it Phys. Rev.} {\bf D 19} (1979) 3080;
H. D. Politzer, {\it Nucl. Phys.} {\bf B117} (1976) 397.



\bibitem{pittori2} G. Martinelli et al., Rome group, {\it Nucl. Phys.} {\bf B445} (1995) 81, 
e-print hep-lat/9411010;
G. Martinelli, S. Petrarca, C.T. Sachrajda, and A. Vladikas,
{\it Phys. Lett.} {\bf B311} (1993) 241, Erratum: {\it ibid} {\bf B317} (1993)
660.

\bibitem{adelaide}
%\cite{Skullerud:2001aw}
%\bibitem{Skullerud:2001aw}
J.~Skullerud, D.~B.~Leinweber and A.~G.~Williams,
%``Nonperturbative improvement and tree-level correction of the quark  propagator,''
Phys.\ Rev.\ D {\bf 64}, 074508 (2001)
[arXiv:hep-lat/0102013];
%%CITATION = HEP-LAT 0102013;%%
%\cite{Bowman:2002bm}
%\bibitem{Bowman:2002bm}
P.~O.~Bowman, U.~M.~Heller and A.~G.~Williams,
%``Lattice quark propagator with staggered quarks in Landau and Laplacian  gauges,''
Phys.\ Rev.\ D {\bf 66}, 014505 (2002)
[arXiv:hep-lat/0203001];
%%CITATION = HEP-LAT 0203001;
%\cite{Bonnet:2002ih}
%\bibitem{Bonnet:2002ih}
F.~D.~Bonnet, P.~O.~Bowman, D.~B.~Leinweber, A.~G.~Williams and J.~b.~Zhang
                  [CSSM Lattice collaboration],
%``Overlap quark propagator in Landau gauge,''
Phys.\ Rev.\ D {\bf 65}, 114503 (2002)
[arXiv:hep-lat/0202003];
%\cite{Zhang:2003fa}
%\bibitem{Zhang:2003fa}
J.~B.~Zhang, P.~O.~Bowman, D.~B.~Leinweber, A.~G.~Williams and F.~D.~Bonnet
                  [CSSM Lattice collaboration],
%``Scaling behavior of the overlap quark propagator in Landau gauge,''
arXiv:hep-lat/0301018.
%%CITATION = HEP-LAT 0301018;%%

%\cite{Cudell:2001ny}
\bibitem{Cudell:2001ny}
J.~R.~Cudell, A.~Le Yaouanc and C.~Pittori,
%``Large pion pole in Z(S)(MOM)/(Z(P)(MOM) from Wilson action data,''
Phys.\ Lett.\ B {\bf 516}, 92 (2001)
[arXiv:hep-lat/0101009];
%%CITATION = HEP-LAT 0101009;%%
%\cite{Cudell:1999kf}
%\bibitem{Cudell:1999kf}
J.~R.~Cudell, A.~Le Yaouanc and C.~Pittori,
%``Pseudoscalar vertex and quark masses,''
Nucl.\ Phys.\ Proc.\ Suppl.\  {\bf 83}, 890 (2000)
[arXiv:hep-lat/9909086];
%%CITATION = HEP-LAT 9909086;%%
%\cite{Cudell:1998ic}
%\bibitem{Cudell:1998ic}
J.~R.~Cudell, A.~Le Yaouanc and C.~Pittori,
%``Pseudoscalar vertex, Goldstone boson and quark masses on the lattice,''
Phys.\ Lett.\ B {\bf 454}, 105 (1999)
[arXiv:hep-lat/9810058].
%%CITATION = HEP-LAT 9810058;%%

%\cite{Boucaud:2000ey}
\bibitem{Boucaud:2000ey}
P.~Boucaud {\it et al.},
%``Lattice calculation of 1/p**2 corrections to alpha(s) and of  Lambda(QCD) in the MOM~ scheme,''
JHEP {\bf 0004}, 006 (2000)
[arXiv:hep-ph/0003020];
%%CITATION = HEP-PH 0003020;%%
%\cite{Boucaud:2000nd}
%\bibitem{Boucaud:2000nd}
P.~Boucaud, A.~Le Yaouanc, J.~P.~Leroy, J.~Micheli, O.~Pene and J.~Rodriguez-Quintero,
%``Consistent OPE description of gluon two point and three point Green  function?,''
Phys.\ Lett.\ B {\bf 493}, 315 (2000)
[arXiv:hep-ph/0008043];
%%CITATION = HEP-PH 0008043;%%
%\cite{Boucaud:2001st}
%\bibitem{Boucaud:2001st}
P.~Boucaud, A.~Le Yaouanc, J.~P.~Leroy, J.~Micheli, O.~Pene and J.~Rodriguez-Quintero,
%``Testing Landau gauge OPE on the lattice with a  condensate,''
Phys.\ Rev.\ D {\bf 63}, 114003 (2001)
[arXiv:hep-ph/0101302];
%%CITATION = HEP-PH 0101302;%%
%\cite{DeSoto:2001qx}
%\bibitem{DeSoto:2001qx}
F.~De Soto and J.~Rodriguez-Quintero,
%``Notes on the determination of the Landau gauge OPE for the asymmetric
three gluon vertex,''
Phys.\ Rev.\ D {\bf 64}, 114003 (2001)
[arXiv:hep-ph/0105063];
%%CITATION = HEP-PH 0105063;%%
%\cite{Boucaud:2002fb}
%\bibitem{Boucaud:2002fb}
P.~Boucaud {\it et al.},
%``OPE and power corrections on the QCD coupling constant,''
arXiv:hep-ph/0205187;
%%CITATION = HEP-PH 0205187;%%
%\cite{Boucaud:2002jt}
%\bibitem{Boucaud:2002jt}
P.~Boucaud {\it et al.},
%``A transparent expression of the A**2-condensate's renormalization,''
Phys.\ Rev.\ D {\bf 67}, 074027 (2003)
[arXiv:hep-ph/0208008].
%%CITATION = HEP-PH 0208008;%%

%\cite{Boucaud:1998bq}
\bibitem{Boucaud:1998bq}
P.~Boucaud, J.~P.~Leroy, J.~Micheli, O.~Pene and C.~Roiesnel,
%``Lattice calculation of alpha(s) in momentum scheme,''
JHEP {\bf 9810}, 017 (1998)
[arXiv:hep-ph/9810322].
%%CITATION = HEP-PH 9810322;%%

%\cite{Becirevic:1999uc}
\bibitem{Becirevic:1999uc}
D.~Becirevic, P.~Boucaud, J.~P.~Leroy, J.~Micheli, O.~Pene, J.~Rodriguez-Quintero and C.~Roiesnel,
%``Asymptotic behaviour of the gluon propagator from lattice {QCD},''
Phys.\ Rev.\ D {\bf 60}, 094509 (1999)
[arXiv:hep-ph/9903364].
%%CITATION = HEP-PH 9903364;%%

%\cite{Becirevic:1999hj}
\bibitem{Becirevic:1999hj}
D.~Becirevic, P.~Boucaud, J.~P.~Leroy, J.~Micheli, O.~Pene, J.~Rodriguez-Quintero and C.~Roiesnel,
%``Asymptotic scaling of the gluon propagator on the lattice,''
Phys.\ Rev.\ D {\bf 61}, 114508 (2000)
[arXiv:hep-ph/9910204];
%%CITATION = HEP-PH 9910204.%%

%\cite{Becirevic:1999kb}
\bibitem{Becirevic:1999kb}
D.~Becirevic, V.~Gimenez, V.~Lubicz and G.~Martinelli,
%``Light quark masses from lattice quark propagators at large momenta,''
Phys.\ Rev.\ D {\bf 61}, 114507 (2000)
[arXiv:hep-lat/9909082];
%%CITATION = HEP-LAT 9909082;%%
%\cite{Becirevic:1999rv}
%\bibitem{Becirevic:1999rv}
D.~Becirevic, V.~Lubicz, G.~Martinelli and M.~Testa,
%``Quark masses and renormalization constants from quark propagator and
3-point functions,''
Nucl.\ Phys.\ Proc.\ Suppl.\  {\bf 83}, 863 (2000)
[arXiv:hep-lat/9909039].
%%CITATION = HEP-LAT 9909039;%%


%\bibitem{A2bare} Ph. Boucaud {\it et al.}, hep-ph/0208008.

%\cite{Chetyrkin:1999pq}
\bibitem{Chetyrkin:1999pq}
K.~G.~Chetyrkin and A.~Retey,
%``Renormalization and running of quark mass and field in the
%  regularization invariant and MS-bar schemes at three and four loops,''
Nucl.\ Phys.\ B {\bf 583}, 3 (2000)
[arXiv:hep-ph/9910332].
%%CITATION = HEP-PH 9910332;%%

%\cite{Luscher:1996ug}
\bibitem{Luscher:1996ug}
M.~Luscher, S.~Sint, R.~Sommer, P.~Weisz and U.~Wolff,
%``Non-perturbative O(a) improvement of lattice QCD,''
Nucl.\ Phys.\ B {\bf 491}, 323 (1997)
[arXiv:hep-lat/9609035].
%%CITATION = HEP-LAT 9609035;%%

%\cite{Neuberger:1997fp}
\bibitem{Neuberger:1997fp}
H.~Neuberger,
%``Exactly massless quarks on the lattice,''
Phys.\ Lett.\ B {\bf 417}, 141 (1998)
[arXiv:hep-lat/9707022];
%%CITATION = HEP-LAT 9707022;%%
%\cite{Neuberger:1998wv}
%\bibitem{Neuberger:1998wv}
H.~Neuberger,
%``More about exactly massless quarks on the lattice,''
Phys.\ Lett.\ B {\bf 427}, 353 (1998)
[arXiv:hep-lat/9801031];
%%CITATION = HEP-LAT 9801031;%%
%\cite{Hernandez:1998et}
%\bibitem{Hernandez:1998et}
P.~Hernandez, K.~Jansen and M.~Luscher,
%``Locality properties of Neuberger's lattice Dirac operator,''
Nucl.\ Phys.\ B {\bf 552}, 363 (1999)
[arXiv:hep-lat/9808010].
%%CITATION = HEP-LAT 9808010;%%
%\cite{Niedermayer:1998bi}
%\bibitem{Niedermayer:1998bi}
F.~Niedermayer,
%``Exact chiral symmetry, topological charge and related topics,''
Nucl.\ Phys.\ Proc.\ Suppl.\  {\bf 73}, 105 (1999)
[arXiv:hep-lat/9810026].
%%CITATION = HEP-LAT 9810026;%%

%\cite{Capitani:1999uz}
\bibitem{Capitani:1999uz}
S.~Capitani, M.~Gockeler, R.~Horsley, P.~E.~Rakow and G.~Schierholz,
%``Operator improvement for Ginsparg-Wilson fermions,''
Phys.\ Lett.\ B {\bf 468}, 150 (1999)
[arXiv:hep-lat/9908029].
%%CITATION = HEP-LAT 9908029;%%

%\cite{Leinweber:1998im}
\bibitem{Leinweber:1998im}
D.~B.~Leinweber, J.~I.~Skullerud, A.~G.~Williams and C.~Parrinello  [UKQCD
                  collaboration],
%``Gluon propagator in the infrared region,'' DEMOCRATIE
Phys.\ Rev.\ D {\bf 58}, 031501 (1998)
[arXiv:hep-lat/9803015].
%%CITATION = HEP-LAT 9803015;%%

%\cite{Capitani:1998mq}
\bibitem{Capitani:1998mq}
S.~Capitani, M.~Luscher, R.~Sommer and H.~Wittig  [ALPHA Collaboration],
%``Non-perturbative quark mass renormalization in quenched lattice QCD,''
Nucl.\ Phys.\ B {\bf 544}, 669 (1999)
[arXiv:hep-lat/9810063].
%%CITATION = HEP-LAT 9810063;%%


\end{thebibliography}
}
\end{document}